\documentclass[11pt, a4paper]{article}
\usepackage[subrefformat=parens,labelformat=parens]{subfig}  
\usepackage{jinstpub}
\usepackage[english]{babel}
\usepackage[utf8x]{inputenc}
\usepackage[T1]{fontenc}

\usepackage{graphicx}
\usepackage{bm}
\usepackage{latexsym}
\usepackage{epsf}
\usepackage{rotating}
\usepackage{epsfig,graphics,rotate,color}
\usepackage{wrapfig}
\usepackage{gensymb}
\usepackage{amssymb}
\usepackage{amsmath}
\usepackage{amsfonts}
\usepackage{array,hhline,dcolumn}
\usepackage[normalem]{ulem}
\usepackage{color}
\usepackage{hyperref}
\hypersetup{
    colorlinks=true,
    linkcolor=blue,
    filecolor=magenta,
    urlcolor=blue,
    breaklinks=true,
    citecolor=blue,
}
\usepackage{placeins}
\usepackage{mwe}
\usepackage{natbib}
\usepackage{comment}

\usepackage{titlesec}
\titlespacing*{\section}{0pt}{1.1\baselineskip}{\baselineskip}

\begin{document}

\title{A deep-learning based raw waveform region-of-interest finder for the liquid argon time projection chamber}

\collaboration{The ArgoNeuT Collaboration}

\author[a]{R.~Acciarri}

\author[a]{B.~Baller}

\author[b]{V.~Basque}

\author[c]{C.~Bromberg}

\author[a]{F.~Cavanna}

\author[c]{D.~Edmunds}

\author[d]{R.S.~Fitzpatrick}

\author[e]{B.~Fleming}

\author[b]{P.~Green}

\author[a]{C.~James}

\author[f]{I.~Lepetic}

\author[g]{X.~Luo}

\author[a]{O.~Palamara}

\author[e]{G.~Scanavini}

\author[h]{M.~Soderberg}

\author[d]{J.~Spitz}

\author[i]{A.M.~Szelc}

\author[j]{L.~Uboldi}

\author[a]{M.H.L.S.~Wang}

\author[a,1]{W.~Wu, \note{Corresponding author.}}
\emailAdd{wwu@fnal.gov}

\author[a]{T.~Yang}

\affiliation[a]{Fermi National Accelerator Lab, Batavia, Illinois 60510, USA}
\affiliation[b]{University of Manchester, Manchester M13 9PL, United Kingdom}
\affiliation[c]{Michigan State University, East Lansing, Michigan 48824, USA}
\affiliation[d]{University of Michigan, Ann Arbor, Michigan 48109, USA}
\affiliation[e]{Yale University, New Haven, Connecticut 06520, USA}
\affiliation[f]{Rutgers University, Piscataway New Jersey 08854, USA}
\affiliation[g]{University of California, Santa Barbara, California, 93106, USA}
\affiliation[h]{Syracuse University, Syracuse, New York 13244, USA}
\affiliation[i]{University of Edinburgh, Edinburgh EH8 9YL, United Kingdom}
\affiliation[j]{CERN, The European Organization for Nuclear Research, 1211 Meyrin, Switzerland}


\abstract{The liquid argon time projection chamber (LArTPC) detector technology has an excellent capability to measure properties of low-energy neutrinos produced by the sun and supernovae and to look for exotic physics at very low energies. In order to achieve those physics goals, it is crucial to identify and reconstruct signals in the waveforms recorded on each TPC wire. In this paper, we report on a novel algorithm based on a one-dimensional convolutional neural network (CNN) to look for the region-of-interest (ROI) in raw waveforms. We test this algorithm using data from the ArgoNeuT experiment in conjunction with an improved noise mitigation procedure and a more realistic data-driven noise model for simulated events. This deep-learning ROI finder shows promising performance in extracting small signals and gives an efficiency approximately twice that of the traditional algorithm in the low energy region of $\sim$0.03-0.1 MeV. This method offers great potential to explore low-energy physics using LArTPCs.}

\keywords{Neutrino detectors; Noble liquid detectors (scintillation, ionization, double-phase);  Time projection Chambers (TPC); Data processing methods}

\maketitle


\section{\label{sec:introduction}Introduction}

The Liquid Argon Time Projection Chamber (LArTPC) detector is a proven technology that has been adopted by many accelerator-based neutrino experiments, including the Short-Baseline Neutrino program at Fermilab~\cite{Antonello:2015lea, Machado:2019oxb} and DUNE~\cite{Abi:2020wmh}. It offers millimeter-scale spatial resolution and excellent calorimetric capabilities in the detection of particles traversing the liquid argon and the measurement of their properties.

Understanding and optimizing the signal and noise discrimination capabilities of LArTPCs is crucial in performing charge reconstruction and, ultimately, for achieving a wide range of physics goals. This is especially critical for low-energy physics, such as low-energy neutrino cross-section measurements~\cite{Formaggio:2013kya}, the study of Michel electrons~\cite{Acciarri:2017sjy},  MeV-scale photons~\cite{Acciarri:2018myr}, solar neutrinos in the $\sim$1 MeV range and core-collapse supernova neutrinos in the $\sim$10 MeV range~\cite{Abi:2020wmh, Acciarri:2018myr}. There are also new physics scenarios at low energies, such as millicharged particles, which can be studied in LArTPCs~\cite{Acciarri:2019jly}. The threshold for extracting small signals such as these is largely determined by the signal-to-noise ratio (SNR). The ArgoNeuT experiment, with its good SNR, has already demonstrated the ability to reconstruct activity at the MeV-scale in a LArTPC~\cite{Acciarri:2018myr}. It is important, however, to continue pushing the limits to achieve even lower thresholds in the detection of low-energy interactions, to make a broader range of exciting physics analyses accessible.

The LArTPC technology provides many advantages with its fine-grained images of neutrino events as well as the wealth of detailed information that can be extracted from the data through automated event reconstruction. However, the capability for detecting low-energy activity has received relatively less attention. In reality, LArTPCs excel in this regard, due to the $\sim$23.6 eV mean energy to ionize an electron in liquid argon, the high ionization electron collection efficiency, and the low level of noise achievable in modern electronics readouts~\cite{Castiglioni:2020tsu}. Beyond this fundamental capability, the threshold for detecting low-energy activity depends mainly on the signal processing algorithms used in LArTPC event reconstruction. These include noise mitigation and the detection and localization of signals in the raw waveforms.

Traditionally, the detection of the presence of signals in raw wire waveforms is based on an over-threshold algorithm that selects signal candidates with pulse heights above a predefined threshold. This method has the disadvantage of discarding true signals below certain energies. In this paper, we introduce a novel deep-learning approach based on the application of a simple one-dimensional convolutional neural network (1D-CNN) to the task of finding regions-of-interest (ROIs) in minimally processed LArTPC waveforms, as described in Ref.~\cite{Uboldi:2021jyj}. Deep learning techniques are widely used in high energy physics and play a significant role in the reconstruction of neutrino interactions. However, most algorithms rely on two-dimensional images as inputs to classify the neutrino interactions~\cite{Abi:2020xvt, Abratenko:2020pbp, Acciarri:2016ryt}. The 1D-CNN ROI finder we describe here can be applied to raw wire waveforms prior to any high-level event reconstruction, thereby preserving the potential for maximizing signal detection efficiency in the initial stages of data analysis, which is absolutely essential for achieving the overall high efficiency required in low-energy physics studies. Since it does not rely on the artificially imposed cuts used in traditional over-threshold algorithms, it has the potential to extend sensitivities to regions below these cuts. Its use can, therefore, substantially enhance the ability to study low-energy physics in LArTPC experiments, as well as help us determine the threshold for detecting low-energy activity in LArTPCs.

ArgoNeuT is a small LArTPC placed 100 m underground in the Neutrinos at the Main Injector (NuMI) beamline at Fermilab just upstream of the MINOS near detector~\cite{Adamson:2011qu}. It has dimensions of $40 \times 47 \times 90$ cm$^3$ [vertical, drift, horizontal (beam)] with a volume of 170 L. The electric field inside the TPC along the drift direction is 481 V/cm. There are two readout wire planes of 240 wires each (the induction and collection planes) angled at $\pm60$ degrees to the beam direction with a plane spacing and wire pitch of 4 mm. Each wire channel is sampled every 198 ns with 2048 time samples (``time ticks")/trigger, for a total readout window of 405.5 $\mu$s. Triggering for the readout window is determined by the NuMI beam spill rate of 0.5 Hz. ArgoNeuT collected neutrino and antineutrino events from September, 2009 through February, 2010. A more detailed description of the ArgoNeuT detector and its operations can be found in Ref.~\cite{Anderson:2012vc}.

This paper is organized as follows: Section~\ref{sec:signal_noise} describes the signal and noise characteristics of raw LArTPC waveforms, followed by the noise mitigation procedure and the data-driven noise model for noise simulation; Section~\ref{sec:cnnroi} describes the 1D-CNN ROI finder for signal and noise discrimination; Section~\ref{sec:results} provides the results from the application of the 1D-CNN ROI finder to ArgoNeuT data and the comparison of its performance with that of the traditional over-threshold algorithm; and finally, Section~\ref{sec:conclusion} presents our conclusions.


\section{\label{sec:signal_noise}Signal and Noise in Raw LArTPC Waveforms}

In LArTPC detectors, the shape of the raw signal waveform is determined by how the charge signal is formed. As the ionizing electrons drift towards the wire planes under the influence of the external electric field, they pass through the wires of the induction plane before finally being collected by the wires on the collection plane. This leads to induction wire signals that are usually bipolar and collection wire signals that are usually unipolar~\cite{Adams:2018dra}.

\begin{figure}[hb!]
\centering
\includegraphics[width=0.48\textwidth, height=0.3\textwidth, trim = 0.1cm 0.1cm 0.1cm 0.0cm, clip]{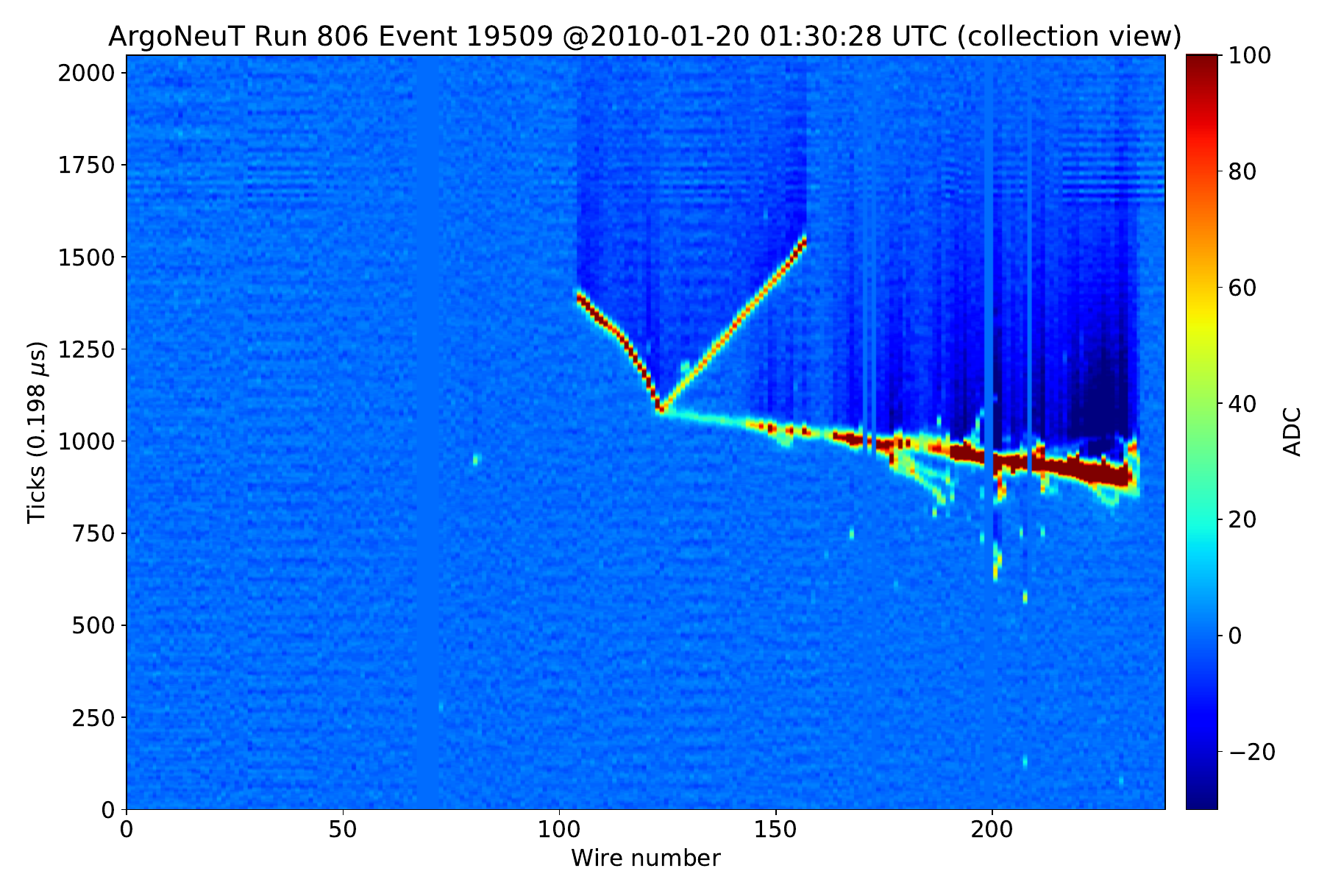}
\includegraphics[width=0.48\textwidth, height=0.3\textwidth, trim = 0.2cm 0.1cm 0.1cm 0.1cm, clip]{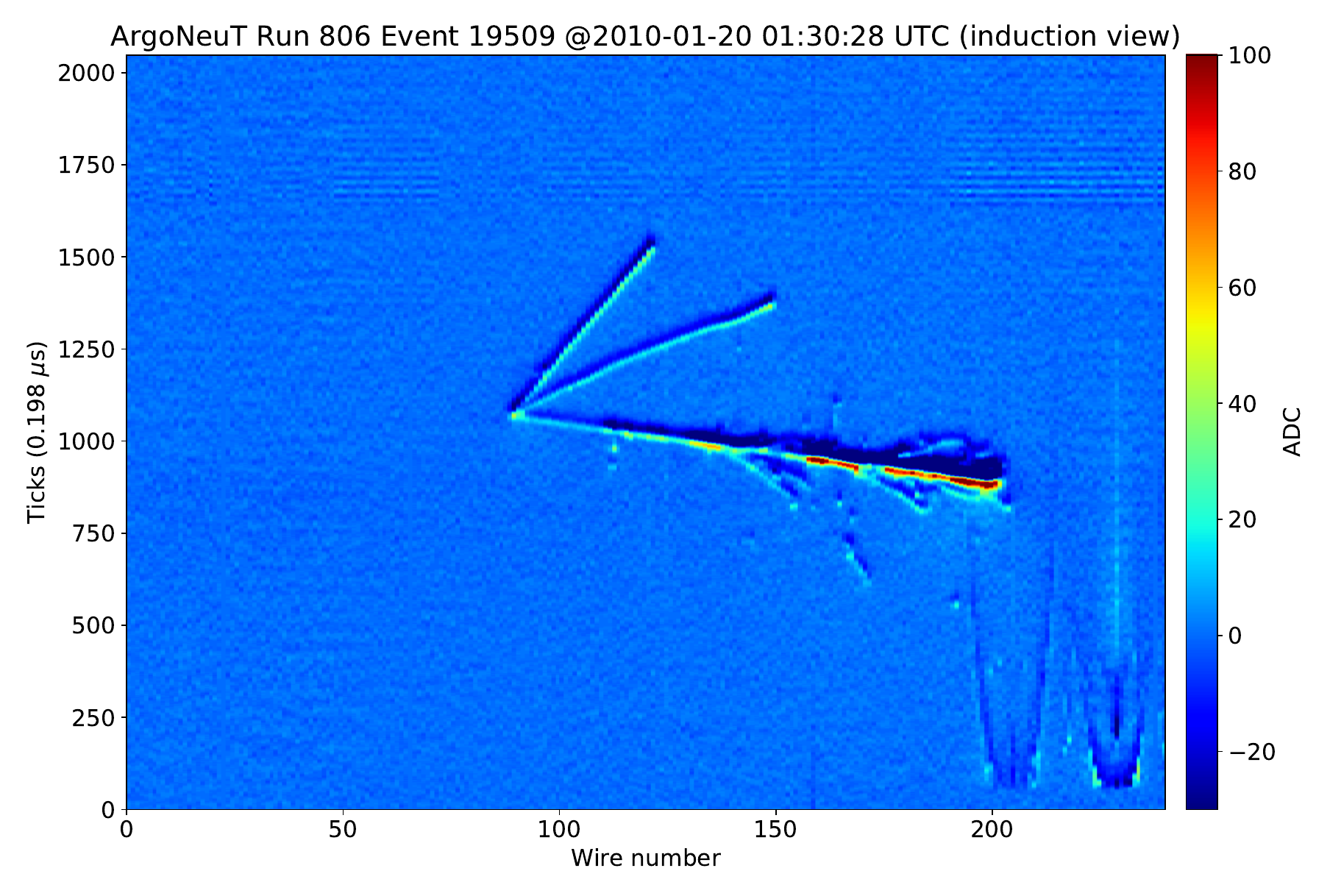}
  \caption{Event display of an electron neutrino interaction candidate in ArgoNeuT data (raw): right view for induction plane and left view for collection plane. The horizontal axis indicates wire number; the vertical axis indicates time ticks; color scale represents the charge measured in ADC counts.}
\label{fig:display_19509_raw}
\end{figure}

Figure~\ref{fig:display_19509_raw} shows an event display of an electron neutrino interaction candidate in ArgoNeuT data based on raw waveforms. Although the signal regions are quite distinct due to the ArgoNeuT detector's good SNR, several noise components are still visible. First is the negative tail (undershoot), shown as the dark blue bands above the signal regions on the collection plane in Figure~\ref{fig:display_19509_raw}. Those tails originate from the capacitive coupling discharge in the ADCs. Second is the coherent noise found across neighboring wire channels on each plane at the same time tick, as shown in Figure~\ref{fig:display_19509_raw}, which is mainly due to power supply line noise, digital noise from readout electronics common to channels occupying the same board or nearby boards, or some external interference. Last is the external noise contribution that only affects the induction plane occasionally, as shown in the bottom right corner of the induction plane view in Figure~\ref{fig:display_19509_raw}. This is related to the charge collected by the bias voltage distribution cards in the liquid argon, which can be released back into the medium. 

\subsection{Noise Mitigation}
\label{subsec:noise_mitigation}

The negative tail and coherent noise components can cause problems for charge reconstruction and need to be removed before further signal and noise discrimination. The noise caused by the charge released from the bias voltage distribution cards is not considered here, since it occurs occasionally and only on the induction plane.

The tails are related to the ADC capacitive coupling discharge. In each TPC wire channel, the amplifier and ADC are AC-coupled through a high-pass RC filter, whose time constant is different for the induction and collection planes. Since signals produced by charged tracks usually occur on short time scales, this AC coupling implies the observed signals will be followed by long tails of opposite sign, whose areas cancel those of the initial signals. This effect is pronounced in the collection plane, but negligible in the induction plane because of the bipolar nature of its signal shape. We make use of an adaptive baseline subtraction method in this paper to deal with the problem in ArgoNeuT. Each TPC wire channel on the collection plane is divided into 64-tick regions and a linear interpolation is performed using the average ADC values in each region and their variances to determine the baseline subtraction applied to the tail regions.

In order to remove the coherent noise, we first determine the wire channel correlation by calculating the Pearson correlation coefficients of noise waveforms between different channels using empty data events with no visible signal. The formula for Pearson's correlation coefficient, $r_{ab}$, is given by:
\begin{equation}
r_{ab} = \frac{\sum a_{i} b_{i} - \sum a_{i} \sum b_{i}}{\sqrt{n \sum a_{i}^{2} - (\sum a_{i})^{2}} \sqrt{n \sum b_{i}^{2} - (\sum b_{i})^{2}}} ~,
\label{eq:correlation}
\end{equation}
where $n = 2048$ is the sample size; $a_i$ and $b_i$ are the ADC values from two different wire channels at time tick $i$.  The wire channels are grouped according to this correlation and the coherent noise removal is performed separately for each time tick by subtracting the median ADC values of all channels in a group from the ADC value of each channel in that group. Although the wire channel correlation exhibits a slight time dependence related to the running conditions, the performance of the electronics, or occasional external interference, the impact on coherent noise removal is small.
 
\begin{figure}[htbp]
\centering
\includegraphics[width=0.48\textwidth, height=0.3\textwidth, trim = 0.1cm 0.1cm 0.1cm 0.0cm, clip]{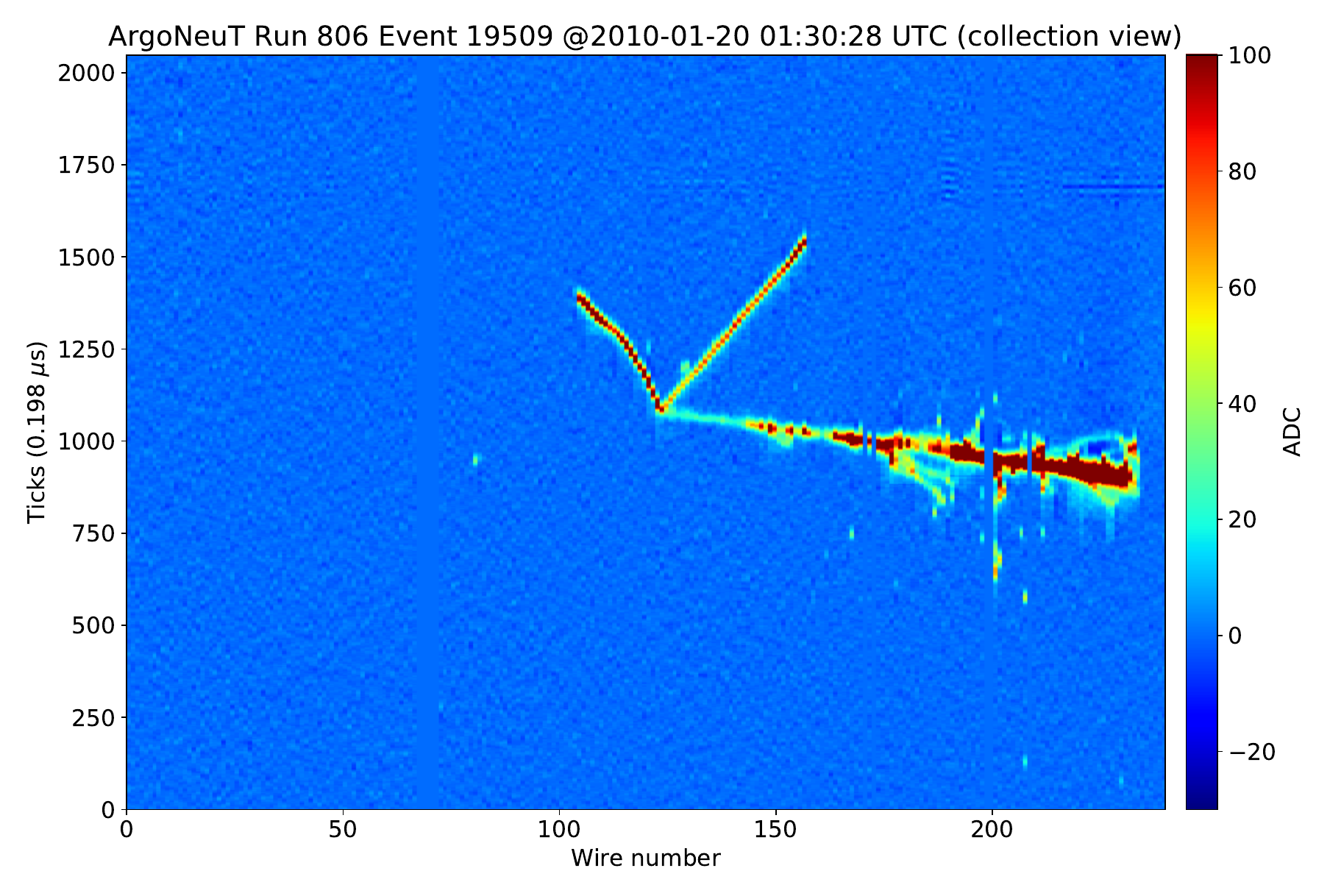}
\includegraphics[width=0.48\textwidth, height=0.3\textwidth, trim = 0.2cm 0.0cm 0.1cm 0.1cm, clip]{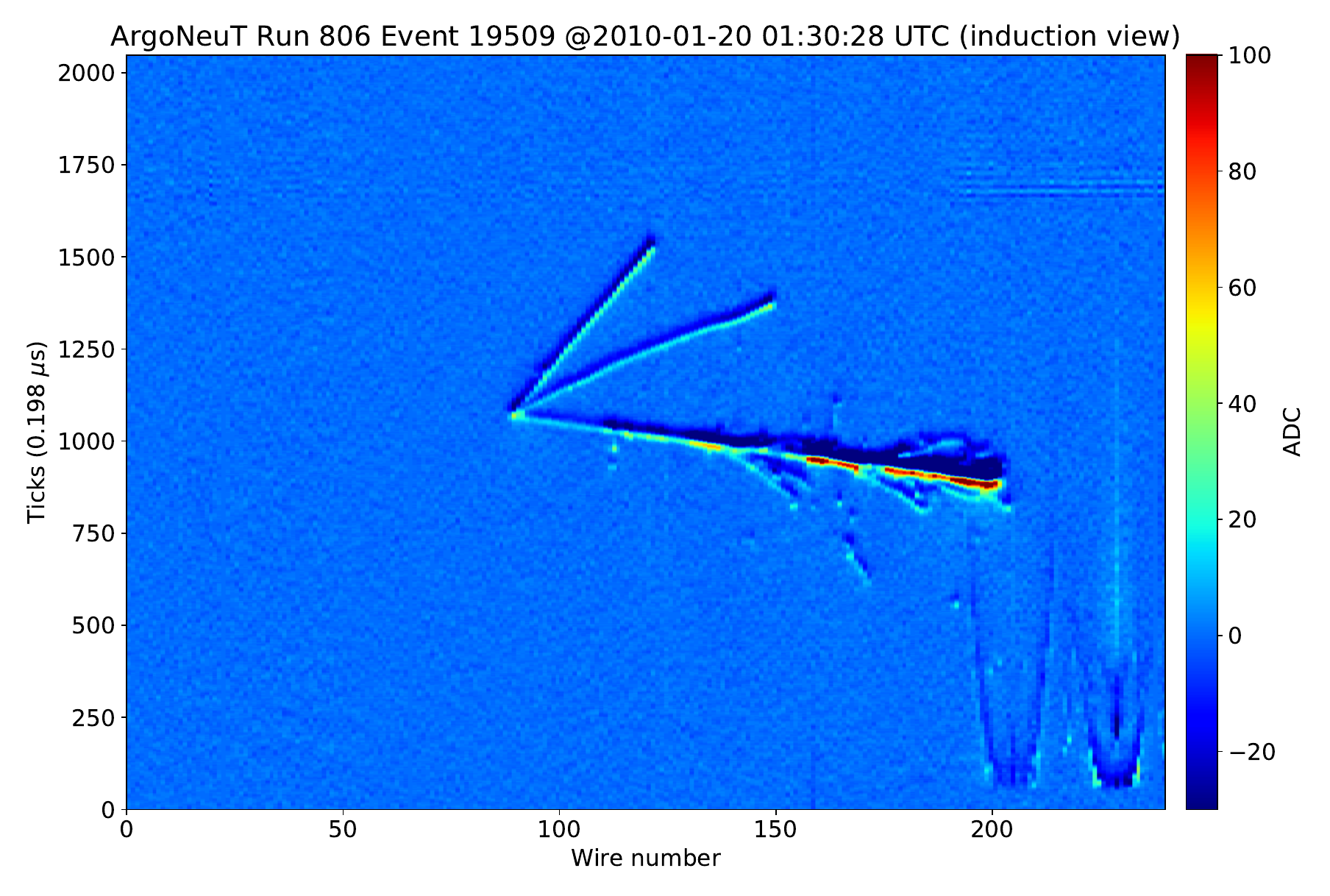}
\caption{Event display after tail and coherent noise removal for the event shown in Figure~\ref{fig:display_19509_raw}.}
\label{fig:display_19509_noiseremoval}
\end{figure}

The tail and coherent noise removal steps outlined above minimize unwanted artifacts from the data that can complicate the task of  signal and noise discrimination in the next stage. They are applied to all ArgoNeuT data used in this work.  However, the removal of coherent noise is not applied to simulated events, since this contribution is not included in the simulations. Figure~\ref{fig:display_19509_noiseremoval} shows the display after tail and coherent noise removal for the event shown in Figure~\ref{fig:display_19509_raw}.

\subsection{Data Driven Noise Model}
\label{subsec:ddn}
In order to understand the noise features better as well as improve the ArgoNeuT noise simulation, we study the noise frequency distribution and develop a data driven noise (DDN) model. We use noise waveforms from selected empty data events and perform a Fast Fourier Transform (FFT) on them. Figure~\ref{fig:ddn_mean} shows the profiled noise frequency distribution on the induction plane, where the error bar in each bin represents the standard deviation (RMS) of the magnitude of the corresponding frequency component. The spikes in the spectrum are due to the remnant coherent noise that is not completely removed.

We account for both the mean value of each noise frequency component and the fluctuation around that mean in ArgoNeuT data with the DDN model as follows. First, we describe the mean of the noise frequency component with the modified exponential function given below:
\begin{eqnarray}
p_{0}\times e^{-0.5\times \left( \frac{x-p_{1}}{p_{2}} \right)^{2}} \times \left( \frac{p_{3}}{x + p_{4}} + 1 \right)  + p_{5} + e^{-p_{6} \times \left( x - p_{7} \right)}~,
\label{eq:ddn_mean}
\end{eqnarray}
where $x$ represents frequency and {$p_{n}$, with $n=0\textup{--}7$, are the parameters determined by fitting this expression to the means of each noise frequency component in ArgoNeuT data for each plane separately. The results are very similar between the two planes and the fitted result for the induction plane is shown in Figure~\ref{fig:ddn_mean}. The mean-normalized magnitude $y$ of each frequency component follows a Poisson-like distribution. We choose to parameterize the fluctuation with a weighted Poisson function given by:
\begin{equation}
\frac{q_{0} \times \left( \frac{q_{1}}{q_{2}} \right)^{\frac{y}{q_{2}}} \times e^{-\frac{q_{1}}{q_{2}}}}{\Gamma\left(\frac{y}{q_{2}} + 1\right)}~,
\label{eq:ddn_rms}
\end{equation}
where $\Gamma(z) = \int_{0}^{\infty} t^{z-1}e^{-t}dt$ is the Gamma function, and the parameters $q_{n}$, with $n=0\textup{--}2$, are determined from a fit to the mean-normalized magnitude of the frequency components, as shown in Figure~\ref{fig:ddn_rms} for the induction plane. Equation~(\ref{eq:ddn_rms}) works very well for almost all frequency bins on both planes. The mean of the noise frequency component can vary with wire length, which is also observed in MicroBooNE \cite{MicroBooNE:2018swd}. For ArgoNeuT, this effect is less than about $5\%$. The mean-normalized magnitude $y$ of each frequency component does not change with wire length.  All simulation results reported in this paper are based on the DDN model. 

\begin{figure}[htbp]
\centering
\includegraphics[width=0.5\textwidth, trim = 0.2cm 0.1cm 0.1cm 0.0cm, clip]{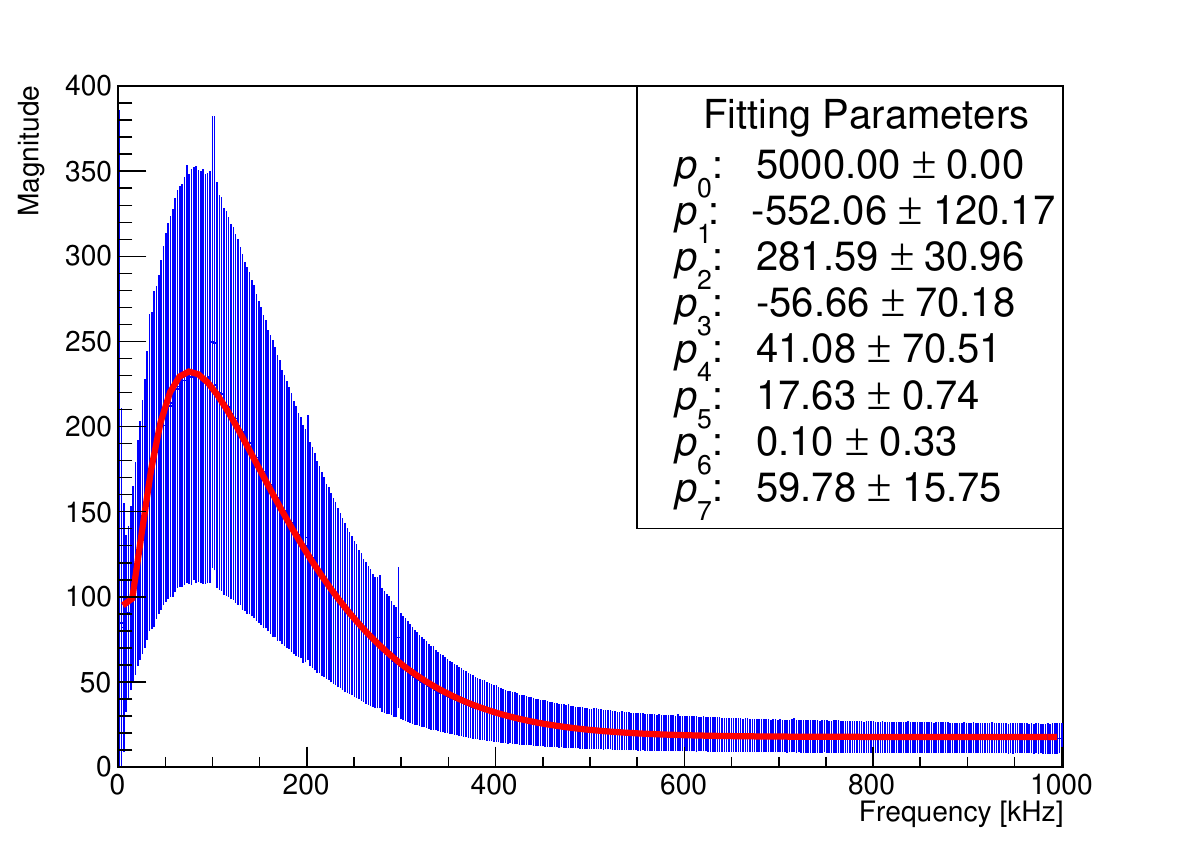}
\caption{Noise frequency on the induction plane after noise removal: error bar in each bin represents the standard deviation (RMS) of the magnitude of the corresponding frequency component; ``spikes" are remnant coherent noise that is not completely removed; red line indicates the fitted result using Eq.~(\ref{eq:ddn_mean}).}
\label{fig:ddn_mean}
\end{figure}

\begin{figure}[htbp]
\centering
\includegraphics[width=0.5\textwidth, trim = 0.0cm 0.0cm 0.0cm 0.0cm, clip]{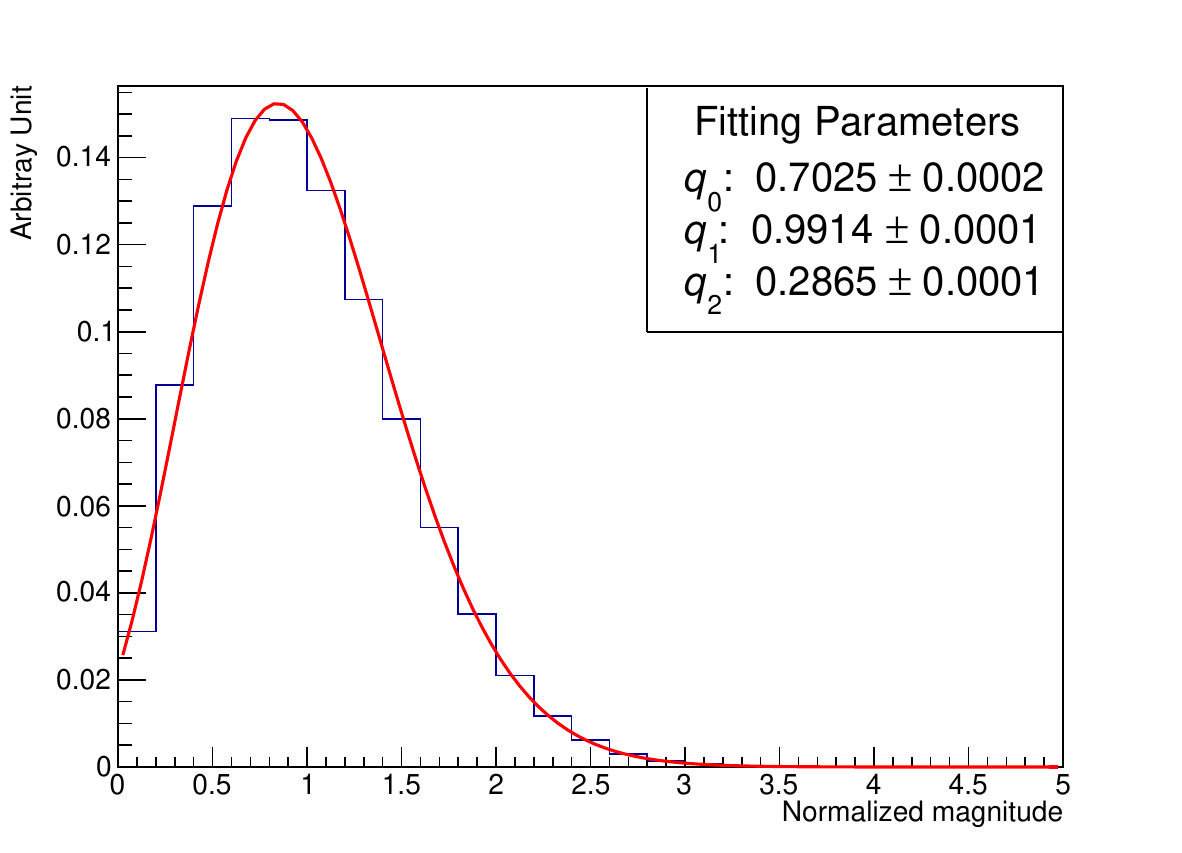}
\caption{Distribution of the mean-normalized magnitude of the noise frequency component at each bin on the induction plane: red line indicates the fitted result using Eq.~(\ref{eq:ddn_rms}).}
\label{fig:ddn_rms}
\end{figure}


\section{\label{sec:cnnroi}A Deep-learning Based  ROI finder}

The ROIs in LArTPC waveforms represent regions that contain ionization electron signals. The deep-learning based waveform ROI finder described in this paper is a one-dimensional convolutional neural network that classifies waveform regions as likely containing signals or not. Unlike most other CNNs used in the classification of neutrino interactions, which rely on two-dimensional images derived from multiple channels in the wire planes, our network looks directly at the one-dimensional waveforms coming from individual channels. Such a 1D-CNN ROI finder can be applied in the earliest stages of reconstruction as a very effective filter to detect ROIs and perform zero-suppression, while maximizing efficiency for usable signals. However, zero-suppression algorithms for a LArTPC with typical wire readouts should be addressed carefully to avoid possible loss of charge.

\subsection{Network Architecture}
\label{subsec:cnn_architecture}

The architecture of the 1D-CNN used in our study is shown in Figure~\ref{fig:cnn_layers}. It is a lightweight network that consists of two convolutional layers, each followed by a max pooling layer, a third convolutional layer followed by a global max pooling layer, and a single output neuron with a sigmoid activation function. The network's kernel size and strides for convolutional operations are chosen in order to be as fast as possible, while achieving the highest accuracy. Overall the network has only 21217 trainable parameters.  The output from the network is a score representing the probability that the input presented to the network contains a signal or not. 

\begin{figure}[htbp]
\centering
\includegraphics[width=0.6\textwidth, trim = 0.0cm 0.4cm 0.0cm -0.1cm, clip]{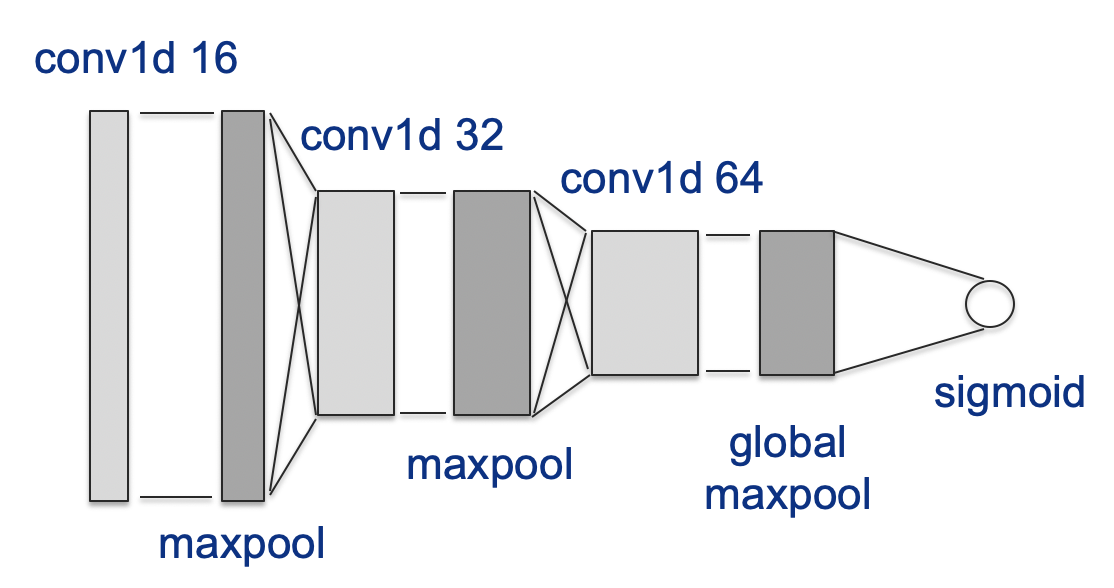}
\caption{Diagram of the 1D-CNN architecture for finding ROIs in raw waveforms.}
\label{fig:cnn_layers}
\end{figure}

The inputs presented to the first layer of the network are partial single-channel LArTPC waveforms with a fixed size of 200 time ticks. For convenience, we will use the word \emph{waveform} in this paper to refer to these partial 200-tick waveforms.

\subsection{Network Training Samples}
\label{subsec:cnn_inputs}

In order for the network to classify these input waveforms properly, they are trained using both classes of waveforms from simulated events -- \emph{signal} referring to those containing signals, and \emph{noise} referring to those that do not.  We define what constitutes a signal within the context of simulated events as a contiguous sequence of time ticks in which there is energy deposition from ionization electrons.  Each signal is characterized by a start time tick $t_{\text{start}}$, end time tick $t_{\text{end}}$, the time tick $t_{\text{max}}$ with the greatest number of ionization electrons, and its value $n_{e}^{\text{max}}$ in that time bin. For simplicity, we use $n_{e}^{\text{max}}$ to represent the size of the signal. Furthermore, if the ionization electrons in a signal originate from more than one parent track, only the parent having the largest contribution to $n_{e}^{\text{max}}$ is associated with the signal. A simulated waveform is labeled as signal if $t_{\text{max}}$ lies within its 200-tick frame.

Noise waveforms are selected randomly from simulated pure noise samples based on the DDN model discussed in Section~\ref{subsec:ddn}. For signal waveforms, since our focus is the low energy region, we simulate the $\beta$-decay of $^{39}$Ar to generate the signal component and select waveforms whose number of electrons range from 200 to 11,000. This allows us to focus on optimizing sensitivity in the sub-MeV region. Waveforms produced by >11,000 ionization electrons can easily be identified as signal waveforms, as shown in Section~\ref{subsec:efficiency}. The noise component of these signal waveforms is based on the same DDN model in the noise waveforms.

Because the induction plane and collection plane have different signal shapes, as discussed in Section~\ref{sec:signal_noise}, separate networks are trained for each plane. The networks in our setup are running on a GPU (consumer Nvidia RTX 2070 SUPER).  For about 1,300,000 windows of 200 time ticks, the inference time is $\sim$1.2 seconds, so around 1 microsecond for each 200 time-tick window. More details of the networks can be found in Ref. \cite{Uboldi:2021jyj}.

\subsection{ROI Reconstruction}
\label{subsec:cnn_roi}

In ArgoNeuT, the full raw waveform from a wire channel has a time window size of 2048 ticks. On the other hand, as mentioned earlier, the inputs to 1D-CNN are 200-tick waveforms.  In order to cover the full waveform, we subdivide it into 14 overlapping 200-tick windows, where each window after the first (whose left edge is aligned with the start of the waveform) is offset from the previous window by a stride length of 150 ticks as shown in Figure~\ref{fig:window_size}.  The exception is the last window which is offset from the previous one by only 48 ticks so that its right edge coincides with the end of the waveform.
This overlap between neighboring windows helps in dealing with signal pulses close to the edge of a window.

\begin{figure}[htbp]
\centering
\includegraphics[scale=0.6, trim = 0.0cm 0.3cm 0.0cm -0.1cm, clip]{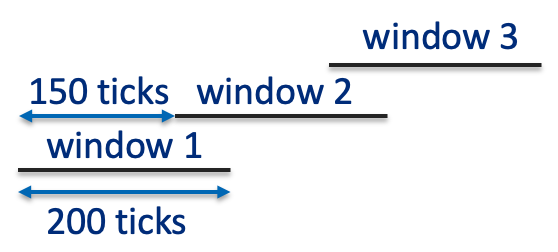}
\caption{Schematic of applying ROI finder with a window size of 200 ticks and a stride size of 150 ticks}
\label{fig:window_size}
\end{figure}

\begin{figure}[htbp]
\centering
\includegraphics[width=0.48\textwidth, trim = 0.0cm 0.0cm 0.0cm 0.0cm, clip]{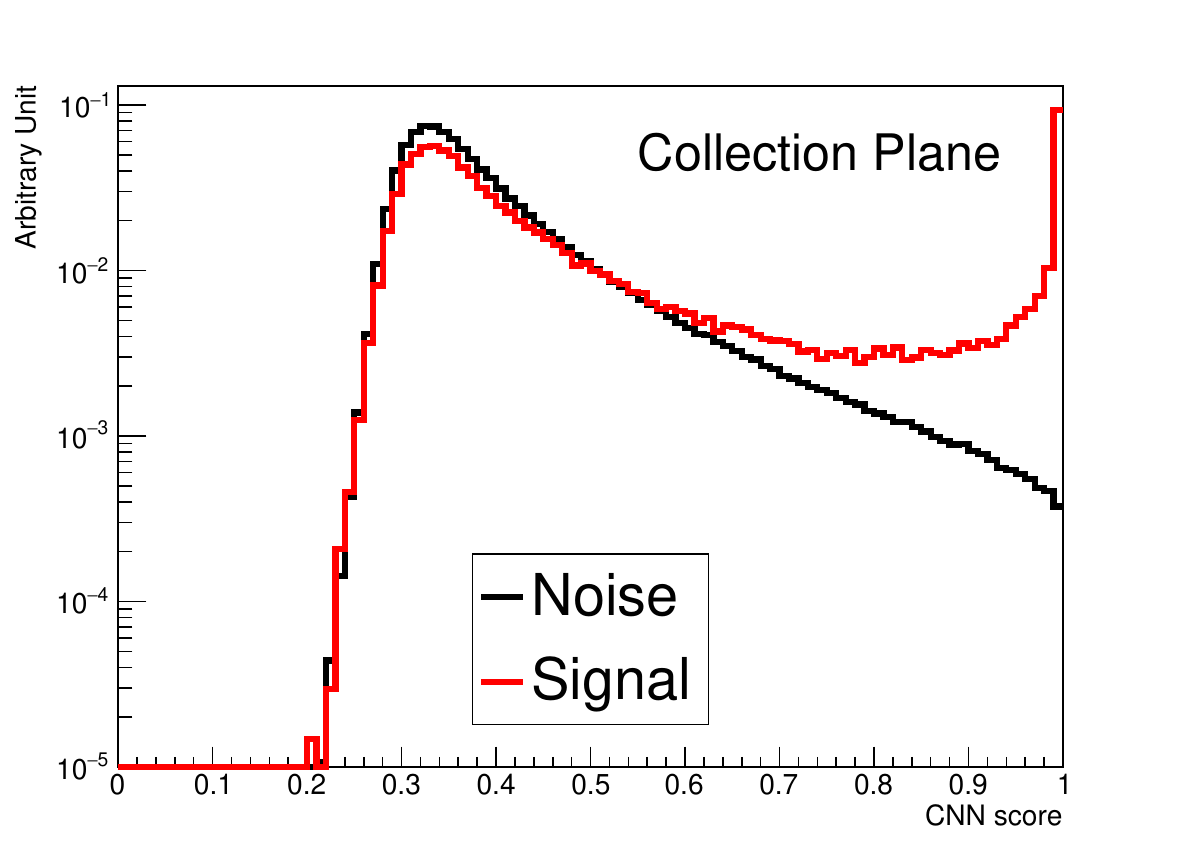}
\includegraphics[width=0.48\textwidth, trim = 0.0cm 0.0cm 0.0cm 0.0cm, clip]{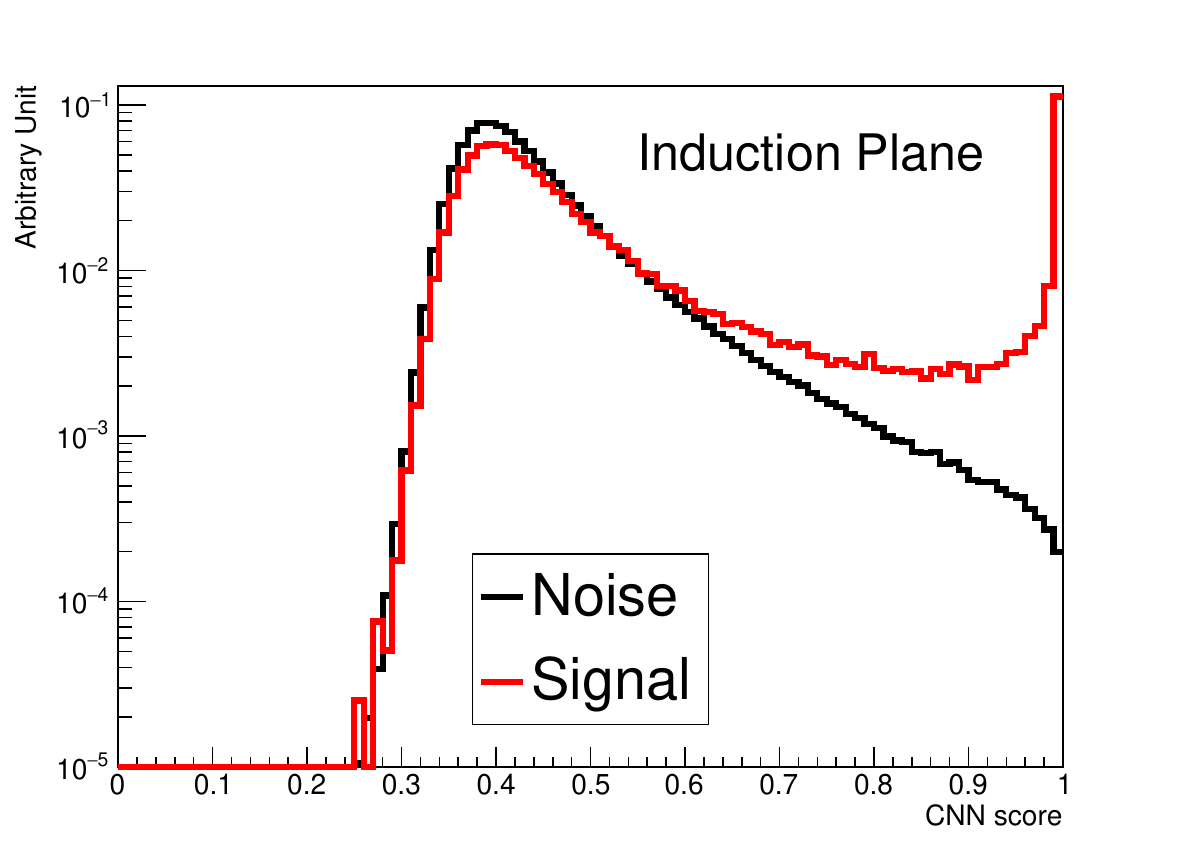}
  \caption{1D-CNN scores for simulated noise and signal wavefoms in the induction plane (right) and the collection plane (left).}
\label{fig:cnn_score}
\end{figure}

Figure~\ref{fig:cnn_score} shows the 1D-CNN scores representing the signal probability for simulated noise and signal waveforms in the induction and collection planes. When applying the 1D-CNN ROI finder to reconstruct ROIs, predefined cuts are chosen to select the signal candidates. If two consecutive windows are both flagged as signal candidates, they are merged into one ROI.

\section{\label{sec:results}Results}

In this section, we present the results of applying our 1D-CNN ROI finder in ArgoNeuT. The ArgoNeuT data we have used is from the antineutrino mode run lasted 4.5 months with $1.25\times 10^{20}$ protons on targets (POT) acquired. For the current analysis, the ArgoNeuT data set and Monte Carlo (MC) simulation generated for the analysis present in Ref. \cite{Acciarri:2018myr} are used. As in Ref. \cite{Acciarri:2018myr}, the selected events are charged-current pion-less events with one muon and up to 1 proton ($\nu_{\mu}$CC 0$\pi$, 0 or 1 proton events) and the MC data set is produced using the FLUKA neutrino interaction generator \cite{Ferrari:2005zk, Battistoni:2009jen, Battistoni:2015epi}.  We choose the simulation of neutrino-argon interactions from FLUKA rather than the GENIE neutrino interaction generator \cite{Andreopoulos:2009rq} because our goal is to prove that our method efficiently reconstruct low-energy signals. Low-energy photons produced in neutrino-argon interactions by the de-excitation of the target nucleus after a neutrino interaction (de-excitation gammas), which are one of the main source of low-energy activities in neutrino interactions, are simulated in FLUKA but not in GENIE. Details of the FLUKA MC simulation event selection can be found at Ref. \cite{Acciarri:2018myr}.

\begin{figure}[htbp]
\centering
\includegraphics[width=0.48\textwidth, height=0.3\textwidth, trim = 0.1cm 0.1cm 0.1cm 0.0cm, clip]{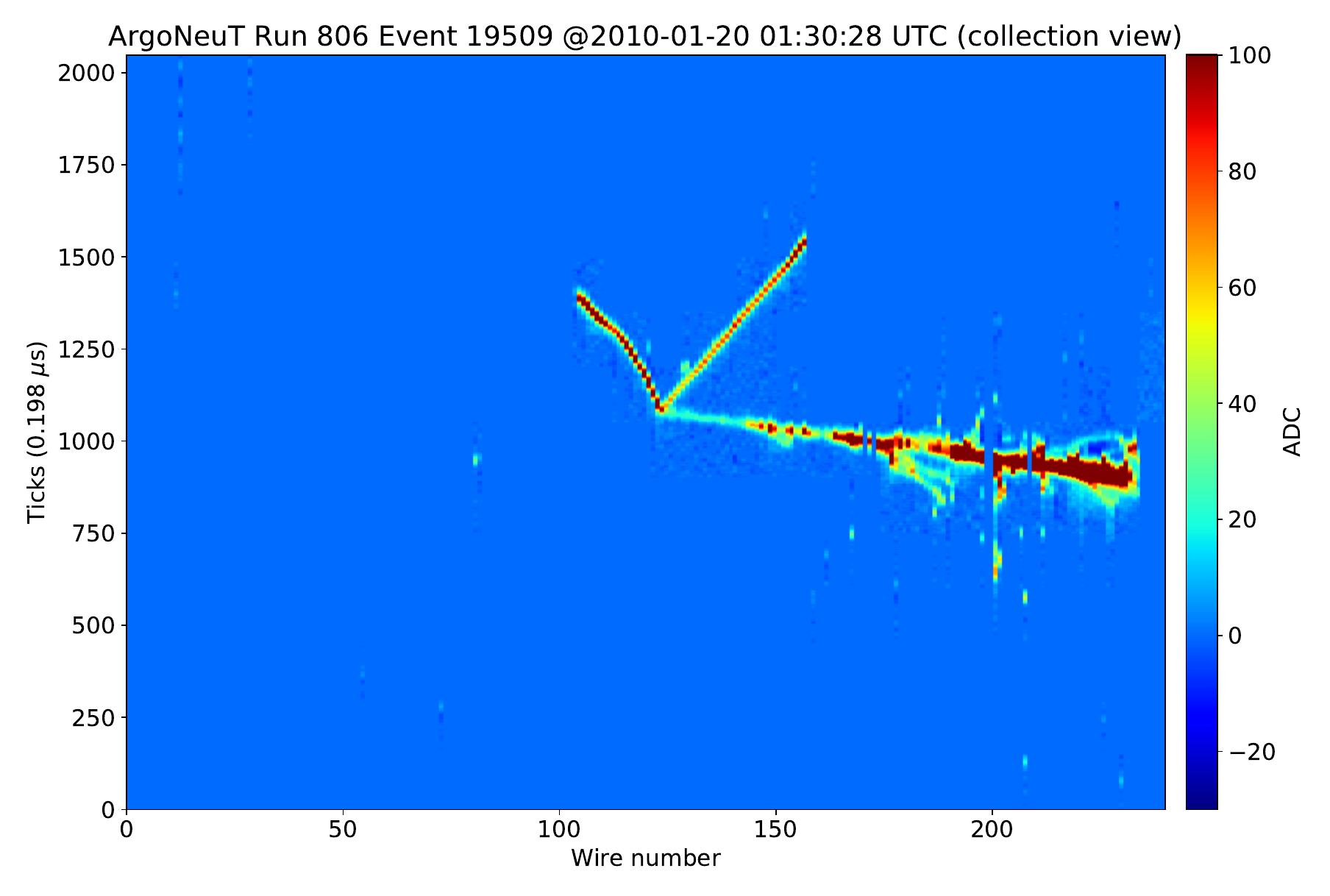}
\includegraphics[width=0.48\textwidth, height=0.3\textwidth, trim = 0.2cm 0.0cm 0.1cm 0.1cm, clip]{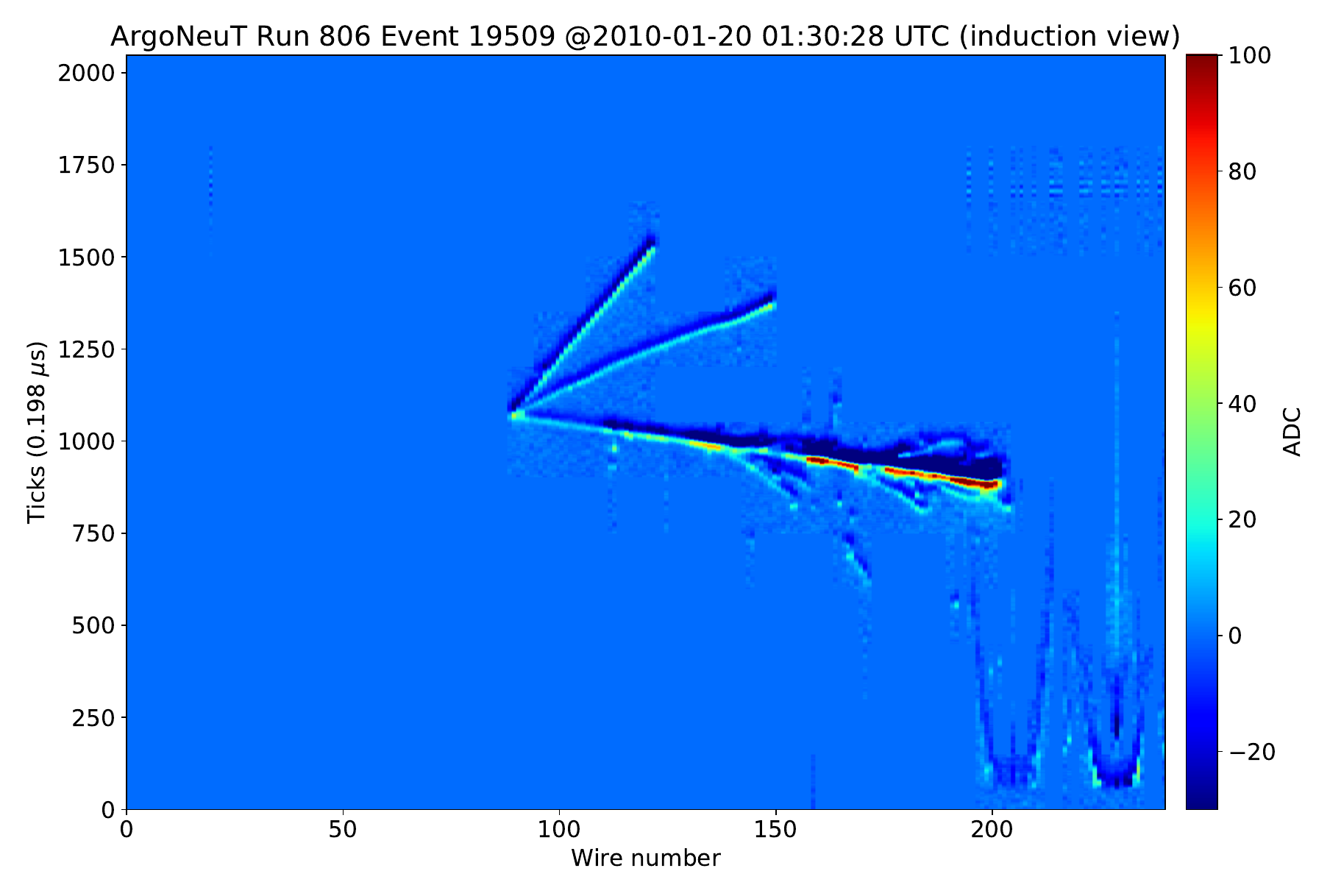}
\caption{Event display after applying the 1D-CNN ROI finder for the event shown in Figure~\ref{fig:display_19509_raw} and Figure~\ref{fig:display_19509_noiseremoval}.}
\label{fig:display_19509_cnnroi}
\end{figure}

In order to establish a baseline relative to which we can evaluate its performance, we also compare our deep learning based method with the traditional over-threshold algorithm mentioned in Section~\ref{sec:introduction}. In the traditional over-threshold algorithm, we use the same strategy described in Section~\ref{subsec:cnn_roi} of subdividing the full 2048-tick waveform into partial overlapping 200-tick waveforms.  However, instead of feeding these partial waveforms to the 1D-CNN for classification, we search for the maximum ADC value within the waveform and flag it as a signal (noise) candidate if it is above (below) a certain threshold. For ArgoNeuT, a threshold cut of 6 ADCs is chosen for both induction and collection planes. This yields noise rejections of $99.95\%$ and $99.94\%$, respectively, on the induction and collection planes. To facilitate comparisons, we require the output of the 1D-CNN ROI finder to be $>0.979$ ($>0.986$) for the induction (collection) plane in order to achieve the same noise rejection as the over-threshold algorithm. Figure~\ref{fig:display_19509_cnnroi} shows the event display for the same event shown in Figure~\ref{fig:display_19509_raw} and ~\ref{fig:display_19509_noiseremoval} after applying the 1D-CNN ROI finder with the above cuts on network scores.

To compare the two waveform ROI finders, we can look at the distributions of  the maximum ADC value within the ROI from the MC simulation, shown in Figure \ref{fig:maxadc_hs} for both ROI finders in the two planes. As shown in Figure \ref{fig:maxadc_hs}, when using the ADC over-threshold ROI finder, any signal below the threshold cut is lost; while the 1D-CNN ROI finder can take advantage of other features, such as the signal shape, to extend sensitivity to signals below this threshold. The relative contributions from various particle types is also shown in Figure \ref{fig:maxadc_hs}, where each ROI is associated with the particle having the largest contribution to $n_{e}^{max}$. As shown in Figure~\ref{fig:maxadc_hs}, most low-energy signal ROIs are from electrons originating from photon interactions (labeled as photon in Figure~\ref{fig:maxadc_hs}). Such photons are mainly from inelastic neutron scattering and de-excitation of the argon nucleus \cite{Acciarri:2018myr}.

\begin{figure}[htbp]
\centering
\includegraphics[width=0.45\textwidth, trim = 0.0cm 0.0cm 0.4cm 0.0cm, clip]{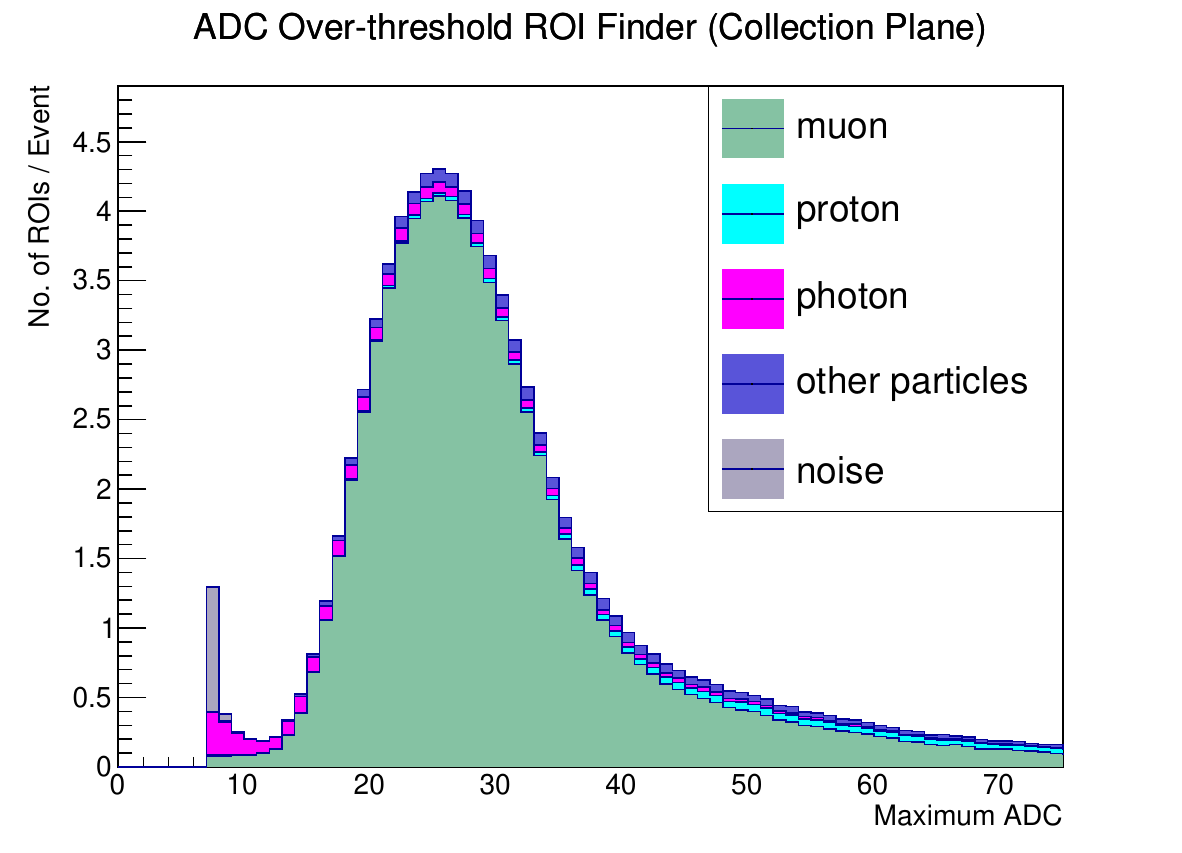}
\includegraphics[width=0.45\textwidth, trim = 0.0cm 0.0cm 0.4cm 0.0cm, clip]{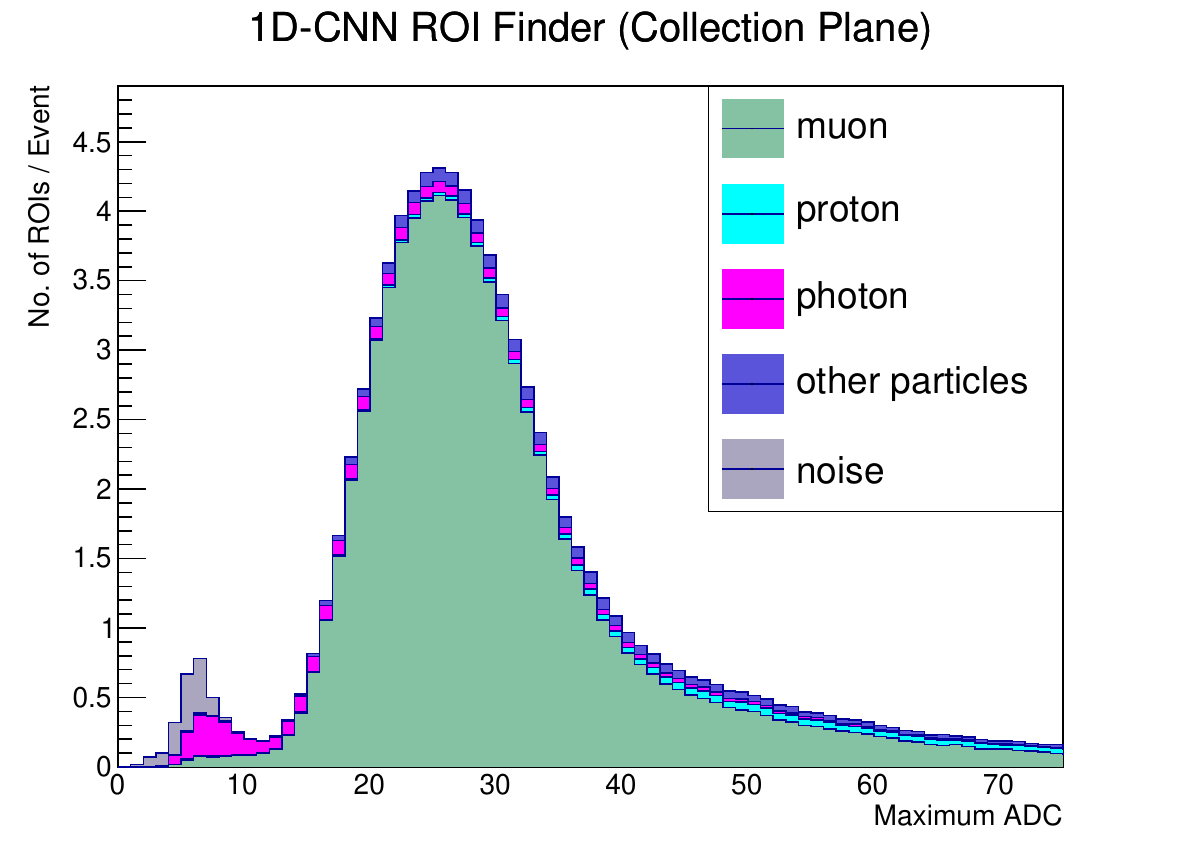}
\includegraphics[width=0.45\textwidth, trim = 0.0cm 0.0cm 0.4cm 0.0cm, clip]{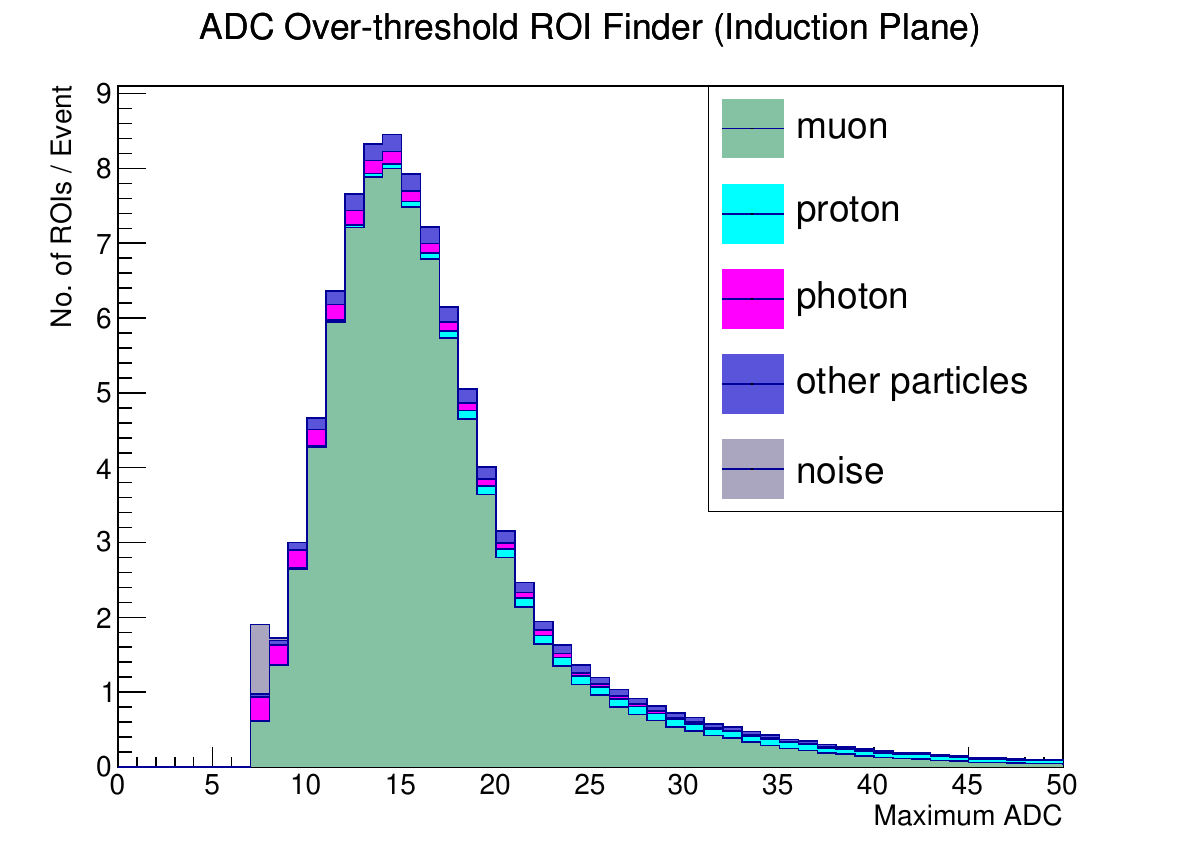}
\includegraphics[width=0.45\textwidth, trim = 0.0cm 0.0cm 0.4cm 0.0cm, clip]{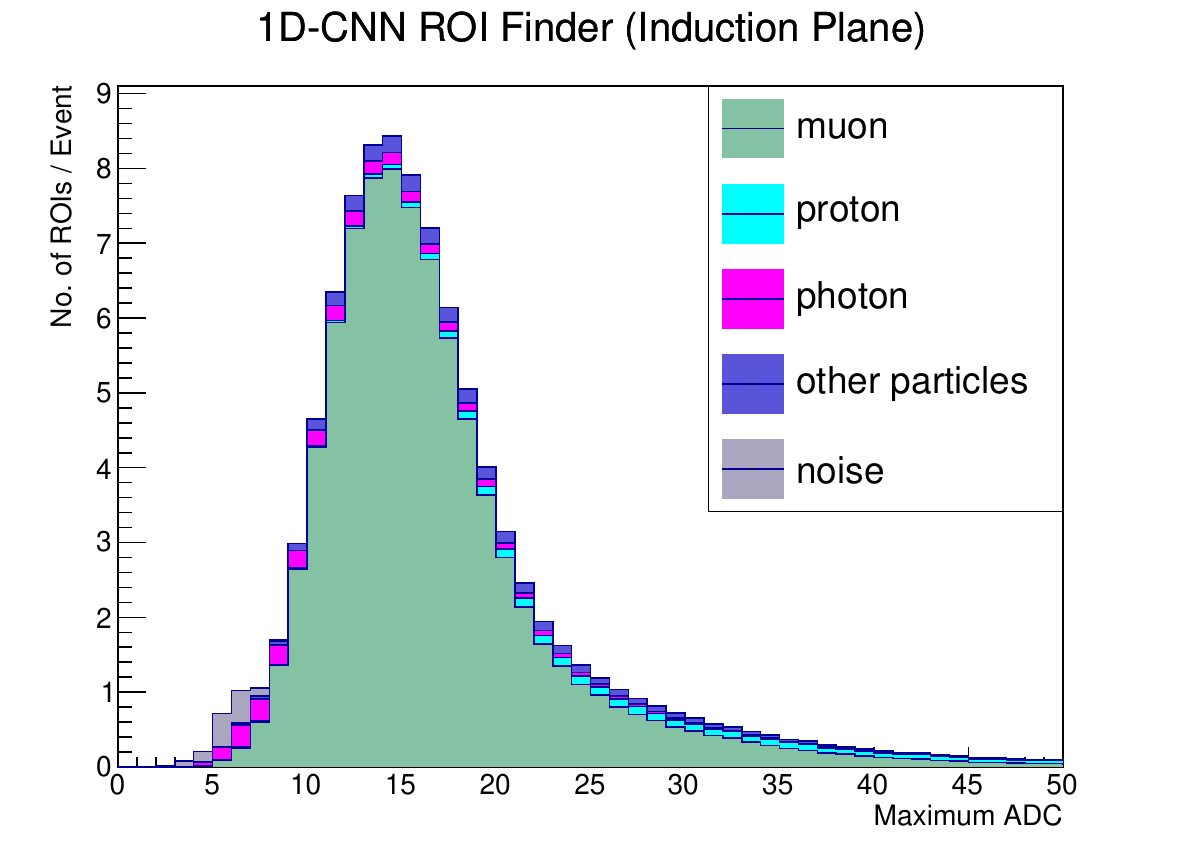}
  \caption{Maximum ADC distributions for the two ROI finders described in the text for both planes using a sample of $\nu_{\mu}$CC 0$\pi$, 0 or 1 proton events from the FLUKA MC simulation. Top and bottom rows are for collection and induction planes, respectively. Left and right columns are for the ADC Over-threshold and 1D-CNN ROI finders, respectively. The ADC over-threshold ROI finder will lose any signal below the threshold cut.}
\label{fig:maxadc_hs}
\end{figure}

\subsection{Comparison of Data and Monte Carlo Simulation}
\label{subsec:datamc}

To compare data and MC simulation, several analysis cuts are applied. We first require that the reconstructed vertex of the selected neutrino events lies within the fiducial volume, defined to be 6 cm from the anode and cathode planes, 6 cm from the top and bottom TPC boundaries, 20 cm from the upstream face of the detector, and 4 cm from the downstream face of the detector. To reduce the impact of ambient gamma ray activity and photons produced by entering neutrons from neutrino interactions occurring upstream of the detector, we skip the first 50 wire channels for each plane. We also skip some noise channels as well as the corner regions which might contain released charge from the bias voltage distribution cards and possible remaining coherent noise, as shown in Figure~\ref{fig:display_19509_noiseremoval}. In addition, to suppress hits originating from above-threshold electronics noise, matching in time of ROI between induction and collection planes is applied. We require the minimum tick difference on the $t_{\text{max}}$ between a ROI from a given wire on one plane and a ROI from possible crossed wires on the other plane should be less than 15 ticks. The electron lifetime corrections are applied to both data and MC simulation and a gain correction is also applied to MC simulation to match data. Such corrections account for the ionization electron loss caused by attachment to impurities in the liquid argon during drift, and for the electronics gain difference between data and simulation, as described in Ref.~\cite{Acciarri:2018ahy}.

\begin{figure}[htbp]
\centering
\includegraphics[width=0.45\textwidth, trim = 0.4cm 0.0cm 0.6cm 0.0cm, clip]{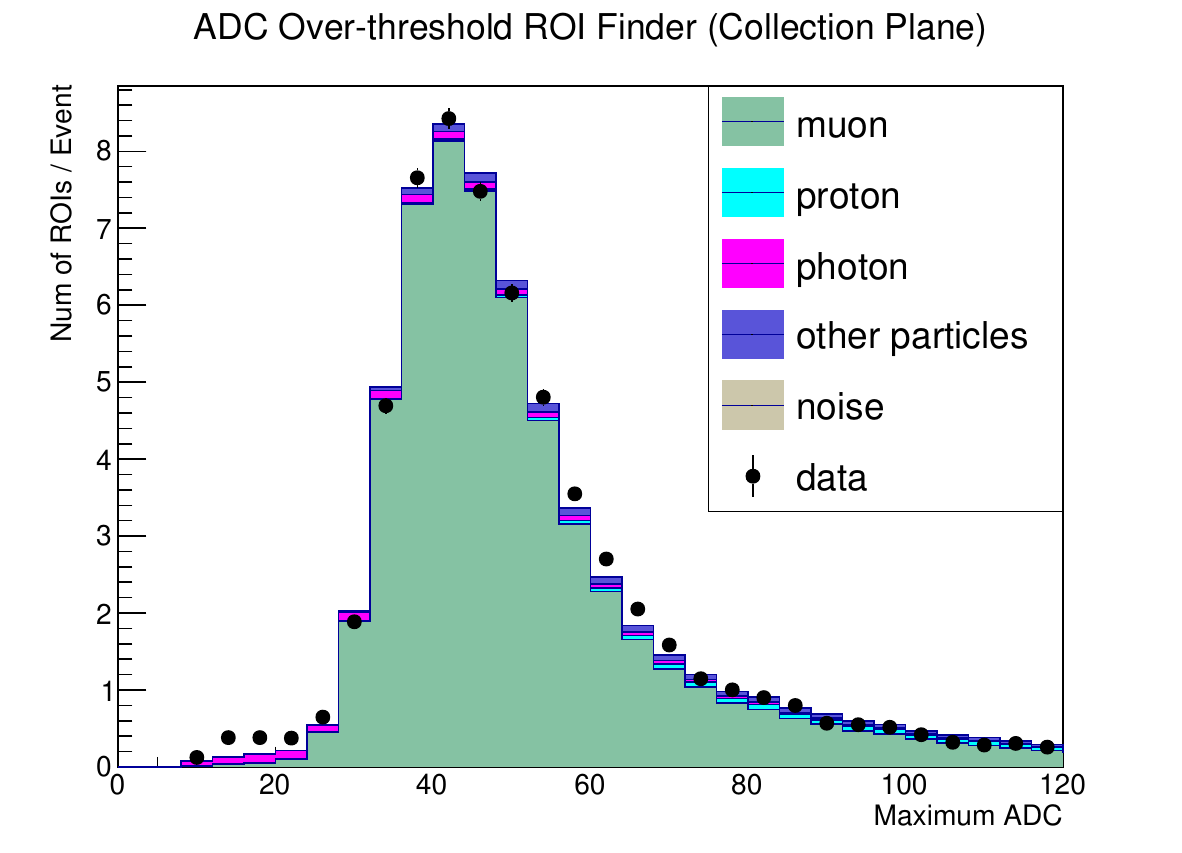}
\includegraphics[width=0.45\textwidth, trim = 0.4cm 0.0cm 0.6cm 0.0cm, clip]{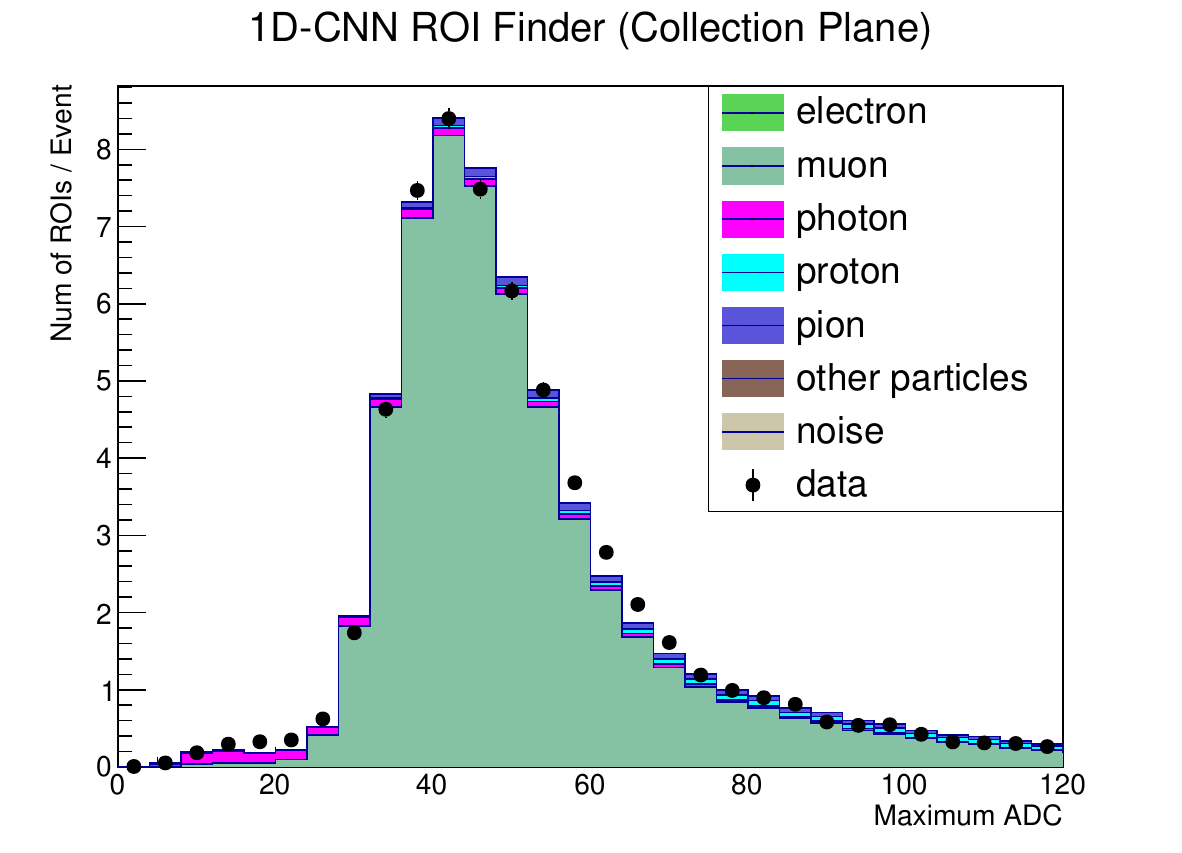}
\includegraphics[width=0.45\textwidth, trim = 0.4cm 0.0cm 0.6cm 0.0cm, clip]{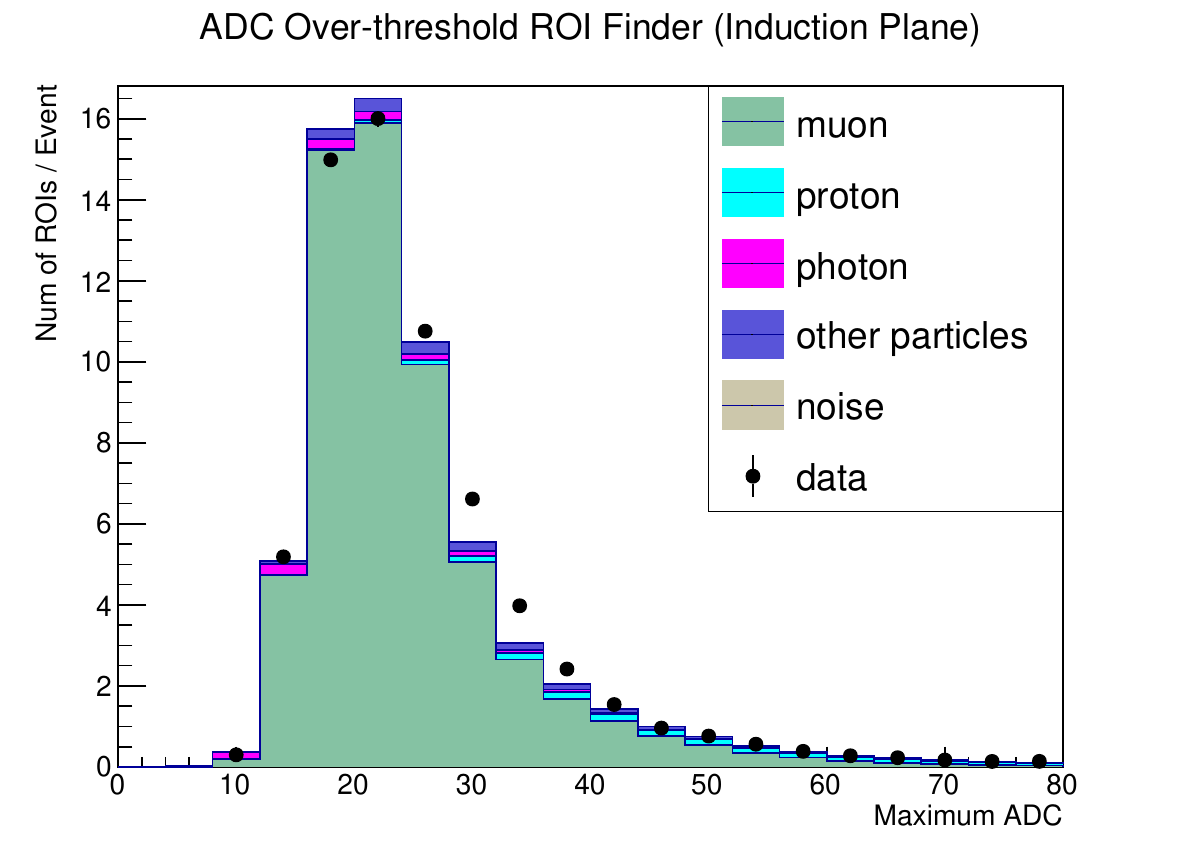}
\includegraphics[width=0.45\textwidth, trim = 0.4cm 0.0cm 0.6cm 0.0cm, clip]{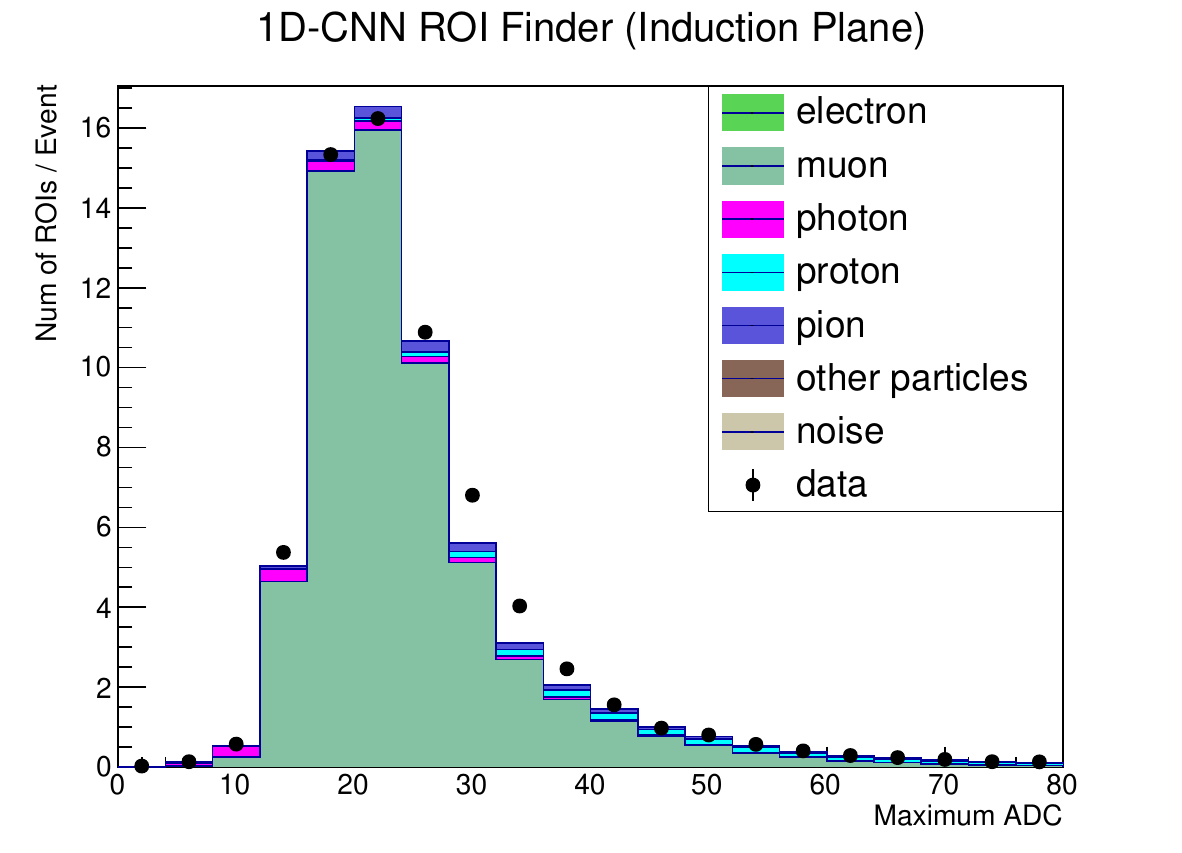}
\caption{Maximum ADC distributions of different ROI finders for both planes for the selected $\nu_{\mu}$CC 0$\pi$, 0 or 1 proton events from data and MC simulation. Top and bottom rows are for the collection and induction planes, respectively. Left and right columns are for the ADC Over-threshold and 1D-CNN ROI finders, respectively.}
\label{fig:maxadc_hs_datamc}
\end{figure}

Figure~\ref{fig:maxadc_hs_datamc} shows the comparison of the maximum ADC distributions normalized based on the number of selected events for the ADC over-threshold ROI finder and the 1D-CNN ROI finder from data and MC simulation. Considering the presence in the data of some posible contributions not included in the MC simulations, there is an overall agreement between data and MC simulation. These are background contributions from electromagnetic activity in the detector originating from neutrino interactions outside the detector's active volume as discussed in Ref.~\cite{Acciarri:2020lhp}, and possible remnants of coherent noise and tails not completely removed by the procedure described in Section \ref{subsec:noise_mitigation}. These known sources contribute to the disagreement between data and MC simulation and have a larger effect on the ADC Over-threshold ROI finder than on the 1D-CNNROI finder at the low-energy region as shown in Figure \ref{fig:maxadc_hs_datamc}. Other than that, the 1D-CNN ROI finder is more suitable for extracting small signals from minimally processed single-channel LArTPC wire waveforms, while the ADC over-threshold ROI finder is unable to recover any signal below the threshold cut.

\subsection{Efficiency}
\label{subsec:efficiency}

We define the waveform ROI efficiency as follows:
\begin{equation}
\text{ROI efficiency}  = \frac{\text{number of signals in ROI}}{\text{number of signals}} ~,
\label{eq:roi_efficiency}
\end{equation}
where signal is defined in Section~\ref{subsec:cnn_inputs}. A signal is considered in a ROI if its $t_\text{max}$ lies within the ROI. If there is more than one signal in the same ROI, only the largest signal is counted. For simplicity, we use the $n_{e}^{\text{max}}$ of the largest signal to represent the signal size of the ROI.

\begin{figure}[htbp]
\centering
\includegraphics[width=0.48\textwidth, trim = 0.0cm 0.0cm 0.0cm 0.0cm, clip]{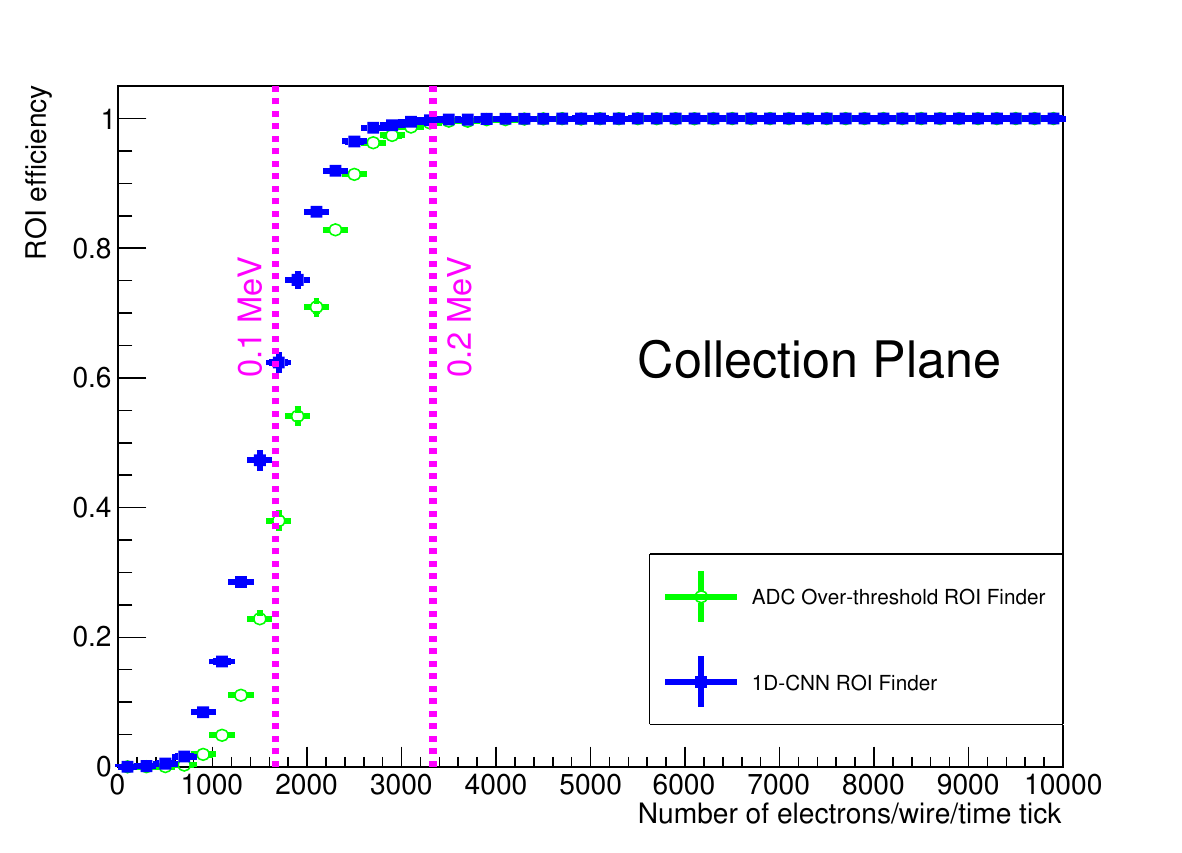}
\includegraphics[width=0.48\textwidth, trim = 0.0cm 0.0cm 0.0cm 0.0cm, clip]{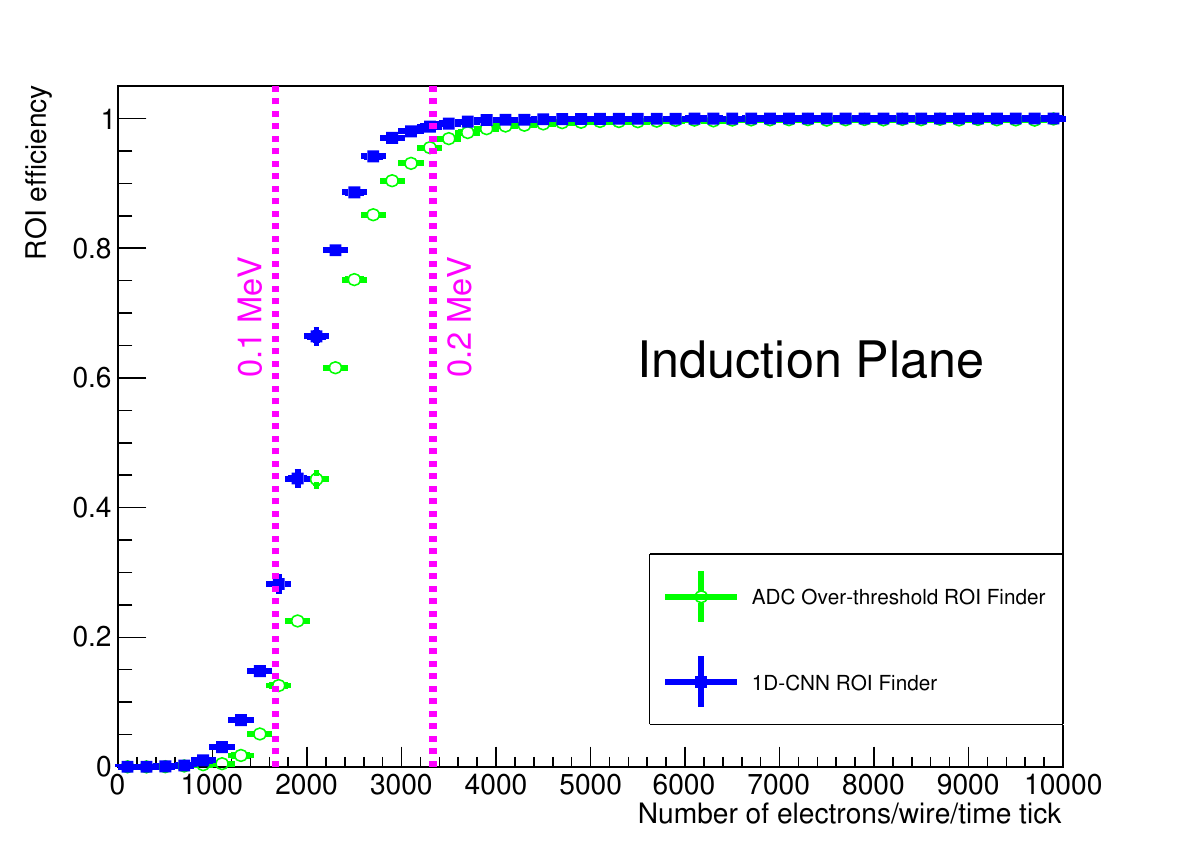}
  \caption{Comparison of the ROI efficiencies as a function of $n_{e}^{\text{max}}$ of the two ROI finders described in the text for the induction plane (right) and the collection plane (left).}
\label{fig:efficiency_roifinders}
\end{figure}

Figure~\ref{fig:efficiency_roifinders} shows the ROI efficiency as a function of $n_{e}^{\text{max}}$ of both the ADC over-threshold ROI finder and the 1D-CNN ROI finder for each of the two planes for the simulated data sample. The cuts described in Section \ref{subsec:datamc} are not applied for the efficiency calculations, because their purpose is to reduce the effects of low-energy activities from upstream/outside TPC and remaining noise features in ArgoNeuT data. {We also show the average deposited energy of 0.1 MeV and 0.2 MeV where the ionization happens in Figure \ref{fig:efficiency_roifinders}. The ionization electrons will drift towards anode planes and only a fraction of them can reach the readout wires and be detected due to the inefficiencies and physics effects such as recombination, attenuation, and diffusion \cite{ArgoNeuT:2013kpa, Acciarri:2018ahy}. Overall, the 1D-CNN ROI finder gives better results than the ADC over-threshold ROI finder on both the induction and collection planes as shown in Figure \ref{fig:efficiency_roifinders}. In the low energy region between $\sim$0.03-0.1 MeV, the efficiency of the 1D-CNN ROI finder is about twice that of the ADC over-threshold ROI finder, which is very promising for exploring low energy physics. Because of the difference in signal shape, both ROI finders show better performance on the collection plane than the induction plane, as shown in Figure~\ref{fig:efficiency_roifinders} and Figure~\ref{fig:efficiency_roiplanes}. As indicated by the dotted vertical magenta lines in the figures, signal waveforms that deposit $\ge0.2$ MeV of energy in $t_{\text{max}}$ are classified as ROIs with $>95$\% efficiency; while those that deposit $<0.1$ MeV of energy in $t_{\text{max}}$ are more challenging to identify as ROIs.

\begin{figure}[htbp]
\centering
\includegraphics[width=0.48\textwidth, trim = 0.0cm 0.0cm 0.0cm 0.0cm, clip]{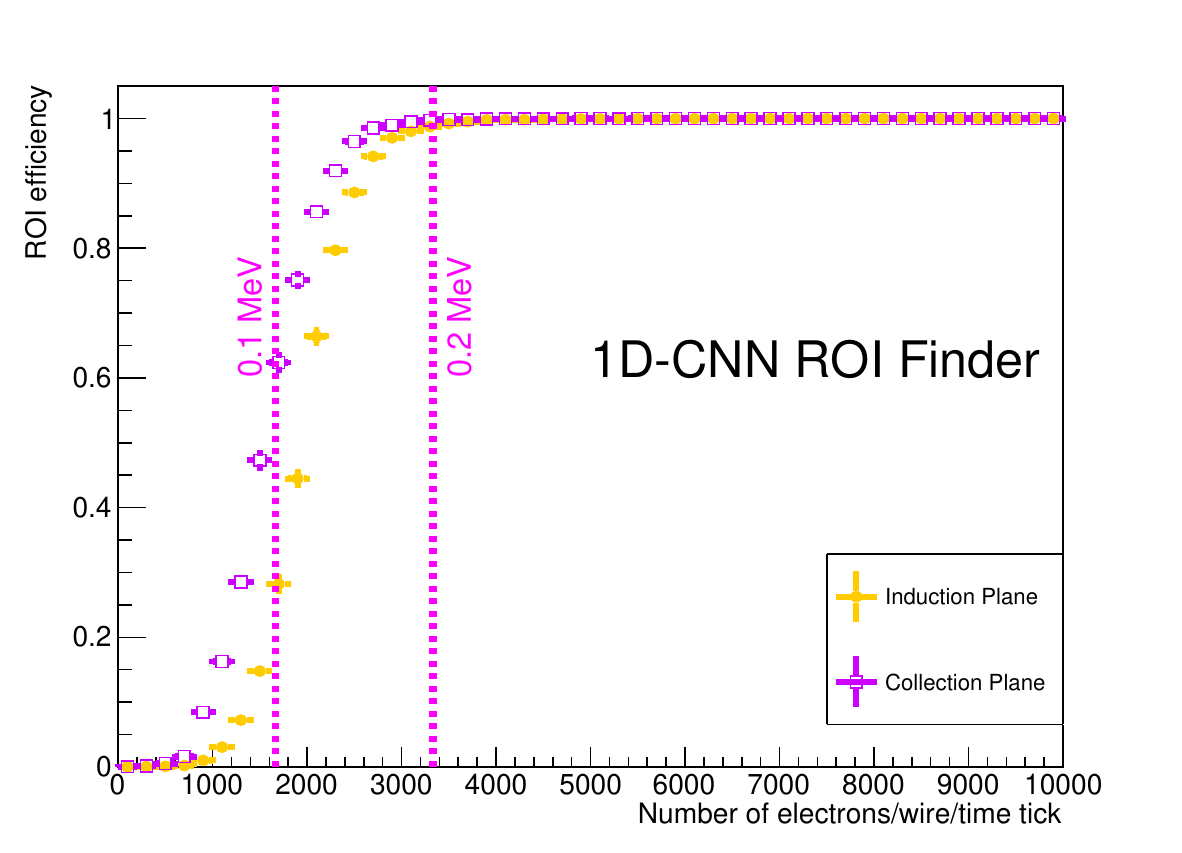}
\caption{ROI efficiencies as a function of $n_{e}^{\text{max}}$ for the 1D-CNN ROI finder in both planes.}
\label{fig:efficiency_roiplanes}
\end{figure}


\section{\label{sec:conclusion}Conclusions}

We have developed a unique deep learning based algorithm for recognizing and localizing signals (ROIs) in the waveforms read out from individual channels of LArTPC detectors. In this paper, we applied this algorithm to the task of ROI finding in minimally processed LArTPC waveforms from the ArgoNeuT detector with very encouraging results. We improved the noise mitigation and built the data-driven noise model in order to explore the algorithm on data. The algorithm employs a 1D-CNN to significantly increase the sensitivity of the ArgoNeuT detector to low energy interactions in liquid argon. The efficiency of our 1D-CNN ROI finder is roughly twice that of a traditional ADC over-threshold algorithm in the very low energy region ($\sim$0.03-0.1 MeV). The ability to recover interesting activity in this region can benefit the very low-energy neutrino physics, such as enhancing our ability to explore solar neutrinos in the $\sim$1 MeV range and core-collapse supernova neutrinos in the $\sim$10 MeV range, both of which are crucial to the physics goals of DUNE~\cite{Abi:2020wmh}. Such a deep-learning based algorithm can easily be optimized and tailored to a LArTPC experiment like DUNE. Because of its potentially higher efficiency and background rejection rate, it can be applied in the initial stages of reconstruction to help reduce data size and speed up data processing. For example, wire channels without ROI candidates in their full output waveforms can be zero-suppressed, reducing the total number of channels written to disk.  Further reduction is possible by storing only the waveform sections representing the ROIs, instead of the full waveforms~\cite{Abratenko:2020hfy}. Because it is based on a fast and lightweight network architecture, it can even be deployed in the upstream stages of a DAQ system as an intelligent filter that can allow the use of more sophisticated trigger algorithms or effectively increase buffer sizes for the storage of longer histories. All these potential applications look very promising indeed for future large-scale LArTPC experiments.

\section*{ACKNOWLEDGEMENT}
This manuscript has been authored by Fermi Research Alliance, LLC under Contract No. DE-AC02-07CH11359 with the U.S. Department of Energy, Office of Science, Office of High Energy Physics. We gratefully acknowledge the collaboration of A. Ferrari, M. Lantz, P. R. Sala and G. Smirnov who provided the FLUKA Monte Carlo simulation used in the analysis. We wish to acknowledge the support of Fermilab, the Department of Energy, and the National Science Foundation in ArgoNeuT’s construction, operation, and data analysis. We also wish to acknowledge the support of the Neutrino Physics Center (NPC) Scholar program at Fermilab.


\bibliographystyle{JHEP}
\bibliography{references}

\end{document}